\documentclass{article}
\usepackage[dvips]{graphicx}
\usepackage{latexsym}

\renewcommand{\arraystretch}{1.3}
\makeatletter
\newdimen\normalarrayskip              
\newdimen\minarrayskip                 
\normalarrayskip\baselineskip \minarrayskip\jot
\newif\ifold             \oldtrue            \def\new{\oldfalse}
\def\arraymode{\ifold\relax\else\displaystyle\fi} 
\def\eqnumphantom{\phantom{(\theequation)}}     
\def\@arrayskip{\ifold\baselineskip\z@\lineskip\z@
     \else
     \baselineskip\minarrayskip\lineskip2\minarrayskip\fi}
\def\@arrayclassz{\ifcase \@lastchclass \@acolampacol \or
\@ampacol \or \or \or \@addamp \or
   \@acolampacol \or \@firstampfalse \@acol \fi
\edef\@preamble{\@preamble
  \ifcase \@chnum
     \hfil$\relax\arraymode\@sharp$\hfil
     \or $\relax\arraymode\@sharp$\hfil
     \or \hfil$\relax\arraymode\@sharp$\fi}}
\def\@array[#1]#2{\setbox\@arstrutbox=\hbox{\vrule
     height\arraystretch \ht\strutbox
     depth\arraystretch \dp\strutbox
     width\z@}\@mkpream{#2}\edef\@preamble{\halign
\noexpand\@halignto
\bgroup \tabskip\z@ \@arstrut \@preamble \tabskip\z@ \cr}%
\let\@startpbox\@@startpbox \let\@endpbox\@@endpbox
  \if #1t\vtop \else \if#1b\vbox \else \vcenter \fi\fi
  \bgroup \let\par\relax
  \let\@sharp##\let\protect\relax
  \@arrayskip\@preamble}
\def\eqnarray{\stepcounter{equation}%
              \let\@currentlabel=\theequation
              \global\@eqnswtrue
              \global\@eqcnt\z@
              \tabskip\@centering
              \let\\=\@eqncr
 \halign to \displaywidth\bgroup
    \eqnumphantom\@eqnsel\hskip\@centering
    $\displaystyle \tabskip\z@ {##}$%
    \global\@eqcnt\@ne \hskip 2\arraycolsep
         $\displaystyle\arraymode{##}$\hfil
    \global\@eqcnt\tw@ \hskip 2\arraycolsep
         $\displaystyle\tabskip\z@{##}$\hfil
         \tabskip\@centering
    &{##}\tabskip\z@\cr}
\begingroup\ifx\undefined\newsymbol \else\def\input#1 {\endgroup}\fi

\catcode`\@=11
\def\marginnote#1{}

\newcount\hour
\newcount\minute
\newtoks\amorpm
\hour=\time\divide\hour by60 \minute=\time{\multiply\hour by60
\global\advance\minute by-\hour}
\edef\standardtime{{\ifnum\hour<12 \global\amorpm={am}%
        \else\global\amorpm={pm}\advance\hour by-12 \fi
        \ifnum\hour=0 \hour=12 \fi
        \number\hour:\ifnum\minute<10 0\fi\number\minute\the\amorpm}}
\edef\militarytime{\number\hour:\ifnum\minute<10 0\fi\number\minute}

\def\draftlabel#1{{\@bsphack\if@filesw {\let\thepage\relax
      \xdef\@gtempa{\write\@auxout{\string
          \newlabel{#1}{{\@currentlabel}{\thepage}}}}}\@gtempa \if@nobreak
    \ifvmode\nobreak\fi\fi\fi\@esphack} \gdef\@eqnlabel{#1}}
    \def\@eqnlabel{}
\def\@vacuum{}
\def\draftmarginnote#1{\marginpar{\raggedright\scriptsize\tt#1}}

\def\draft{
  \oddsidemargin -.5truein
  \def\@oddfoot{\footnotesize \sl preliminary draft \hfil
    \rm\thepage\hfil\sl\today\quad\militarytime}
  \let\@evenfoot\@oddfoot \overfullrule 3pt
    \let\label=\draftlabel
    \let\marginnote=\draftmarginnote
  \def\@eqnnum{(\theequation)\rlap{\kern\marginparsep\tt\@eqnlabel}%
    \global\let\@eqnlabel\@vacuum}

  }

\voffset= - 1.in
\hoffset= - 1.0in
\def\nn{\nonumber}
\def\beq{\begin{equation}}
\def\eeq{\end{equation}}

\def\ba{\beq\new\begin{array}{c}}
\def\ea{\end{array}\eeq}
\def\be{\ba}
\def\ee{\ea}

\newfont{\alef}{msbm10 at 12pt}
\newfont {\goth}{eufm10 at 11pt}
\def\mathbb#1{\hbox{{\alef #1}}}

\let\@@savethanks\thanks
\def\thanks#1{\gdef\thefootnote{\alph{footnote}}\@@savethanks{#1}}

\def\theequation{\arabic{section}.\arabic{equation}}

\title{{\bf "Anomaly" in $n=\infty$ Alday-Maldacena Duality for Wavy Circle} \vspace{.5cm}}
\author{{\bf H. Itoyama}\thanks{E-mail: \
itoyama@sci.osaka-cu.ac.jp}\date{ } \\
{\small {\it Osaka City University, Japan}} \\ \\
{\bf A. Mironov}\footnote{E-mail: \ mironov@itep.ru; mironov@lpi.ru}
\date{ } \\
{\small {\it Lebedev Physics Institute}
and {\it ITEP, Moscow, Russia}}\\ \\
{\bf A. Morozov}\thanks{E-mail: \ morozov@itep.ru}
\date{ } \\ {\small {\it ITEP, Moscow, Russia}}
}

\begin{document}

\maketitle

\vspace{-11.cm}

\begin{center}
\hfill OCU-PHYS 292\\
\hfill FIAN/TD-15/08\\
\hfill ITEP/TH-11/08\\
\end{center}

\vspace{9.0cm}

\begin{abstract}
\noindent If the Alday-Maldacena version of string/gauge duality is
formulated as an equivalence between double loop and area integrals
{\it a la} arXiv:\ 0708.1625, then this pure geometric relation can
be tested for various choices of $n$-side polygons. The simplest
possibility arises at $n=\infty$, with polygon substituted by an
arbitrary continuous curve. If the curve is a circle, the minimal
surface problem is exactly solvable. If it infinitesimally deviates
from a circle, then the duality relation can be studied by expanding
in powers of a small parameter. In the first approximation the
Nambu-Goto (NG) equations can be linearized, and the peculiar NG
Laplacian $\Delta_{NG}= \Delta_0 - {\cal D}^2 + {\cal D}$ plays the
central role. Making use of explicit zero-modes of this operator
(NG-harmonic functions), we investigate the geometric duality in the
lowest orders for small deformations of arbitrary shape lying in the
plane of the original circle. We find a surprisingly strong
dependence of the minimal area on regularization procedure affecting
"the boundary terms" in minimal area. If these terms are totally
omitted, the remaining piece is regularization independent, but
still differs by simple numerical factors like 4 from the
double-loop integral which represents the BDS formula so that we
stop short from the first non-trivial confirmation of the
Alday-Maldacena duality. This confirms the earlier-found discrepancy
for two parallel lines at $n=\infty$, but demonstrates that it
actually affects only a finite number (out of infinitely many) of
parameters in the functional dependence on the shape of the
boundary, and the duality is only {\it slightly} violated, which
allows one to call this violation an {\it anomaly}.
\end{abstract}

\bigskip

\def\thefootnote{\arabic{footnote}}

\newpage


\section{Introduction}
\setcounter{equation}{0}

\subsection{Alday-Maldacena duality}

The Alday-Maldacena version \cite{am1}
of the string-gauge duality \cite{sgd}
is one of the most spectacular new hypotheses of the last year
and it naturally attracts an increasing attention
\cite{amfirst}-\cite{amlast}.
We prefer to formulate it in a pure geometric form \cite{mmt1}:

\paragraph{Conjecture:}
An explicit regularization can be found such that
for any polygon $\Pi$
which is made from $n$ light-like segments
in Minkowski space $R^4_{-+++}$
\be
D_\Pi \equiv
\left. \left(\oint\oint_\Pi
\frac{d\vec y d\vec y\,'}{(\vec y-\vec y\,')^2}\right)
\right|_{\rm regularized}
\ \stackrel{?}{=}\
\left.\Big({\rm Minimal\ Area}\Big)\right|_{\rm regularized}
\equiv A_\Pi
\label{sgd}
\ee
where $A_\Pi$ is (regularized) area of a minimal surface
in the bulk $AdS_5$ space with the metric
\be
ds^2 = \frac{dr^2+d\vec y\,^2}{r^2}, \ \ \
d\vec y\,^2 = -dy_0^2 + dy_1^2 + dy_2^2 + dy_3^2
\label{adsmetric}
\ee
bounded by the polygon $\Pi$ which is located at the
boundary (absolute) of the $AdS_5$ (at $r\to 0$).

\bigskip

\subsection{Comments}

The Alday-Maldacena duality is motivated by considerations
of the planar ($N=\infty$) limit of
${\cal N}=4$ SYM and combines a number of different
hypotheses about the non-perturbative properties of
this theory.
Despite we are going to analyze (\ref{sgd}) as a formal
relation, without direct reference to its physical meaning,
a few remarks are still necessary to clarify the possible
subtleties of the problem.
For more detailed presentation of our understanding of
physical motivation behind (\ref{sgd}) see \cite{mmt1,mmt2,imm,im8}.

\bigskip

{\bf 1.}
$\Pi$ in (\ref{sgd}) is a polygon in the {\it momentum}
space, formed by $n$ null momenta of external gluons.
Therefore $AdS_5$ space at the r.h.s. of (\ref{sgd})
is dual \cite{KT,am1} to the ordinary one in AdS/CFT
correspondence \cite{AdS/CFT}.
Accordingly one needs to distinguish
between conformal $SO(4,2)$ symmetries of the
bulk and momentum $AdS_5$ spaces.

\bigskip

{\bf 2.} The l.h.s. of (\ref{sgd}) looks like a logarithm
of the average of an ordinary {\it Abelian} Wilson loop:
\be
D_\Pi =
\log \left< \exp \left\{i\oint_\Pi A_\mu\!(\vec y)dy^\mu\right\}
\right>_{\rm regularized}
\ee
Eq.(\ref{sgd}) should not be confused with another
well-known conjecture,
\be
A_\Pi \ \stackrel{?}{=} \ \log W_\Pi,
\ee
relating the r.h.s. of (\ref{sgd}) to an average of the
$N=4$ SUSY Wilson loop
\be
W_\Pi = \left< {\rm Tr}
\,P\!\exp \left\{i\oint_\Pi
\Big(A_\mu\!(\vec y)dy^\mu + \phi\, dl\Big)
\right\}\right>_{\rm regularized}
\ee
involving non-Abelian vector fields and scalars and
non-trivial multi-loop diagrams.

\bigskip

{\bf 3.}
The l.h.s. of (\ref{sgd}) is an identical, though non-trivial,
reformulation \cite{am1,dks,bht} of the celebrated BDS conjecture
\cite{bds}, stating that the $n$-gluon MHV amplitude in ${\cal N}=4$
SUSY YM in the planar limit is {\it exactly} equal to the exponential of the
one-loop result, which is in turn reduced to contribution of the
"2me" box diagrams and explicitly expressed through
dilogarithm functions \cite{bddk,yugo}.

\bigskip

{\bf 4.}
The r.h.s. of (\ref{sgd}) can be considered as a version of the
Gross-Mende conjecture \cite{gm} that the high-energy asymptotics
of stringy scattering amplitudes are given by exponentiated
minimal areas in the relevant bulk spaces with appropriate
boundary conditions.
Within the ${\cal N}=4$ SUSY context, one can assume that the statement
is true for all values of external momenta, not obligatory
large, while the ADS/CFT conjecture \cite{AdS/CFT}
identifies the relevant bulk space in this case as
$AdS_5 \times S^5$.

\subsection{Current status of the Alday-Maldacena duality}

The status is somewhat controversial.

All reliable evidence in support of (\ref{sgd}) is
at $n=4$ \cite{am1}
and sometime at $n=5$ \cite{mmt2,dhks1,dhks2}.
Unfortunately, this evidence is not decisive, because at $n=4,5$
explicit expressions are fully determined by the
anomalous Ward identities \cite{koma},
associated with the global conformal invariance of the
problem \cite{dks,dhks1,dhks2,dhks3}.
For $n\geq 6$ this symmetry is too small to unambiguously
constrain the answer, but in this case there is still no clear
way to explicitly evaluate the r.h.s. of (\ref{sgd}).
This Plateau minimal-surface problem is considered
unresolvable (in any explicit form) in flat spaces.
If (\ref{sgd}) was true, this would imply that the situation
is drastically different in $AdS$ space,
since the l.h.s. is an absolutely explicit expression:
the $AdS$ Plateau problem would be {\it exactly solvable},
and this is what makes the Alday-Maldacena hypothesis so
interesting and significant far beyond ${\cal N}=4$ SUSY studies.
Attempts to solve the $AdS$ Plateau problem are
described in \cite{imm,im8}, but they are still far from
being conclusive.

Meanwhile, the counter-arguments {\it against} (\ref{sgd})
are mounting.
Already known ones can be divided into three categories.

Counter-arguments of the first type argue that the BDS conjecture,
which is behind the l.h.s. of (\ref{sgd}), contradicts some other
physically-expected properties of the scattering amplitudes for
$n\geq 6$, like Regge behavior \cite{nasregge}.

The second type of counter-arguments \cite{dhks3}
is based on results of higher-loop calculations of non-Abelian
Wilson average $W_\Pi$.
The claim is that $D_\Pi \neq \log W_\Pi$, so that (\ref{sgd})
comes in contradiction with the usual belief that
$A_\Pi = \log W_\Pi$. This belief is just supported once again by
\cite{bdkrsvv,dhks4}.

The third type \cite{am3,ina} comes from attempts to evaluate
$A_\Pi$ for some special polygons $\Pi$,
when the $AdS$ Plateau problem is simplified.
While in \cite{ina} the boundary conditions are considered which
seem to be inconsistent with the simplest BDS conjecture
(additional restrictions on virtual momenta in the loops
are imposed), the discrepancy found in \cite{am3} can be
eliminated only by an ugly change of regularization,
what signals about a real problem.

All these difficulties look very serious and seem to distract
people from the Alday-Maldacena hypothesis, at least, in its
simplest form (\ref{sgd}).
However, the above-mentioned counter-arguments have a common drawback:
they are too special to show any way out, they can serve only
to rule out formula (\ref{sgd}), but can not explain
how and why it should be modified.
Thus, one needs at this moment is a considerable extension
of the above counter-examples, taking them from particular
selected points in the infinite-dimensional "moduli space"
of all relevant polygons $\Pi$ to at least some vicinities of those:
this can help to get rid of regularization ambiguities
(provided there is only a finite number of possible counterterms)
or to formulate explicit requirements to infinite-parametric
regularization schemes (if one is going to look for a resolution
of emerging problems this way).

\subsection{The goal of this paper:
a perturbative analysis of the smooth $n=\infty$ limit }

In this paper we are going to elaborate on the so far most
constructive counter-example to (\ref{sgd}): the one found in
\cite{am3} for a special rectangular configuration at $n=\infty$.
The specifics of this "smooth $n=\infty$ limit", see s.2.8 of
\cite{imm}, is that the Plateau problem can be reduced from $AdS_5$
to Euclidean $AdS_3$ lying at $y_0=y_3=0$, and $\Pi$ in (\ref{sgd})
becomes an {\it arbitrary} curve in the plane of the complex
variable $z=y_1+iy_2$. One can apply the methods, developed in
\cite{imm,im8}, to solve the Plateau problem, at least, in the from
of power series in the deviations from some exactly-solvable
examples where the role of $\Pi$ is played by two parallel lines or
a circle. This kind of slightly deviating boundary conditions was
called "wavy" in \cite{Sem} (see also \cite{PR}), and, in these
terms, we are going to address the problem of "the wavy circle". The
long-rectangular (actually, the two-parallel-lines) example of
\cite{am3} would correspond to a circle of infinite radius, however,
this large-radius limit is somewhat singular and "wavy rectangular"
requires separate consideration, which is straightforward, but left
beyond the scope of the present paper\footnote{ One could of course
use the results of \cite{PR,Sem}, but since the $\sigma$-model
action was used there instead of the Nambu-Goto one, there are
additional sources of complications.}. In this way we obtain the
l.h.s. and the r.h.s. of (\ref{sgd}) for an infinitely-parametric
family of wavy curves $\Pi$ and thus obtain a significantly wider
information than in the previous considerations.

Our result is somewhat surprising: we confirm that
(\ref{sgd}) is not true, at least, in the most naive
regularization prescription.
However, even for this prescription the two sides of
(\ref{sgd}) are very similar.
Still, they are different, moreover, their global conformal
properties do not coincide.
At the same time, we observe an unexpectedly strong dependence
on the choice of regularization prescription,
what makes the hypothesis formulated in s.1.1,
much more difficult to overturn.

\subsection{The main result of this paper}

Our attempt to confirm relation (\ref{sgd})
for a continuous curve $\Pi=\bar\Pi$, which is
an infinitesimal deformation of a unit circle
in the complex $z$-plane, $z = y_1+iy_2$, with
$y_0=y_3=0$,
{\bf fails}, but in an interesting and puzzling way:
the two sides of (\ref{sgd}) are different, but only
{\it slightly} different.

Namely, if $\Pi$ is an image of the unit circle $|\zeta|^2=1$
under the conformal map
$z = H(\zeta) = \zeta + \sum_{k=0}^\infty h_k\zeta^k$, then
\be
\begin{array}{ccccccccccccccc}
{D_\Pi\over 2\pi} & =& \frac{L}{\lambda} &-&2\pi &-&4\pi\Big[
Q^{(2)}_\Pi &-& Q^{(3,1)}_\Pi &-& Q^{(3,2)}_\Pi\Big] &+& 4\pi
Q^{(4)}_\Pi &+& O(h^5), \label{Dans}
\end{array}
\ee

\be
\begin{array}{ccccccccccccc}
{A_\Pi\over 2\pi} & =& \frac{L}{4\mu} &-& 1& -&{3\over 2}\Big[
Q^{(2)}_\Pi &-&Q^{(3,1)}_\Pi &-& 4Q^{(3,2)}_\Pi \Big]&+& O(h^4)
\label{Aans}
\end{array}
\ee We see the discrepancy between these two expressions: first the
coefficients in front of the brackets differ by a factor of
$\kappa_{\circ}={8\pi\over 3}$, second, one of the structures in the
brackets in $A_\Pi$ differs from those in $D_\Pi$ by a mysterious
integer factor $4$. Thus, only few of infinitely many coefficients
in $h$-expansions are different, still the difference exists even if
regularizations are matched, $\kappa_{\circ}\lambda=4\mu$ and
nonphysical constants $2\pi$ and $1$ are omitted.

Moreover, one could even think that the overall coefficient
$\kappa_{\circ}$ is not a problem at all. However, it is, if one
assumes this coefficient is completely independent on the shape of
$\Pi$. Indeed, in the quadrilateral $n=4$ example
\cite{am1,mmt1,popo}\footnote{In this example, the finite piece of
the double loop integral is $-2(\log s/t)^2$ (see (2.16)-(2.17) and
(2.13) in \cite{mmt1}), while that of the minimal area is $-1/4(\log
s/t)^2$ \cite{am1,popo}, compare with $-1/2(\log s/t)^2$ in the
$\sigma$-model case \cite[(4.26)]{mmt1}.} $\kappa_{\Box}=8$ and,
therefore, $\kappa_{\circ}={8\pi\over 3}={\pi\over 3}\kappa_{\Box}$.
Still, one can imagine a simple dependence of this coefficient only
on the number of corners of $\Pi$ to reproduce this overall
difference ${\pi\over 3}$.

In these formulas $Q^{(p,q)}$ are certain structures of the order
$h^p$:
\be
Q^{(2)}_\Pi = \sum_{k=0}^\infty B_k|h_k|^2, \ \ \ \ \ \ \ \
B_k = \frac{k(k-1)(k-2)}{6},
\ee
\be
Q^{(3)}_\Pi = Q^{(3,1)}_\Pi + Q^{(3,2)}_\Pi =
\frac{1}{2}\sum_{i,j=0}^\infty C_{ij}\Big(h_ih_j\bar h_{i+j-1} +
\bar h_i\bar h_j h_{i+j-1}\Big), \nn \\
C_{ij} = \frac{ij}{6}\Big(i^2+3ij+j^2-6i-6j+7\Big)
\ee
\be
Q^{(3,2)}_\Pi = \frac{1}{2}\sum_{i=0}^\infty C_{ii}
\Big(h_i^2\bar h_{2i-1} + \bar h_i^2 h_{2i-1}\Big),
\ee
while $Q^{(3,1)}_\Pi$ is the sum of off-diagonal terms,
\be
Q^{(3,1)}_\Pi = \sum_{i<j}^\infty C_{ij}
\Big(h_ih_j\bar h_{i+j-1} + \bar h_i\bar h_j h_{i+j-1}\Big)
\ee
Different coefficients in front of $Q^{(3,1)}_\Pi$ and
$Q^{(3,2)}_\Pi$ in $A_\Pi$ imply that the tensor
$C_{ij}^{(A)}=3Q^{(3,1)}_{ij} + 12Q^{(3,2)}_{ij}$,
which would play the role of $C_{ij} = C_{ij}^{(D)}
=Q^{(3,1)}_{ij} + Q^{(3,2)}_{ij}$ in (\ref{Aans}),
is {\it not} just a polynomial in the indices $i,j$.
\be\label{Us}
Q^{(4)}_\Pi = (h_1^2+\bar h_1^2)Q^{(2)}_\Pi
+ \frac{1}{4}\sum_{\stackrel{i,j,k,l=0}{i+j=k+l}}^\infty
U_{ij;kl}h_ih_j\bar h_k\bar h_l +
\frac{1}{6}\sum_{i,j,k=0}^\infty V_{ijk}
\Big(h_ih_jh_k\bar h_{i+j+k-2} + \bar h_i\bar h_j\bar h_k h_{i+j+k-2}
\Big), \nn \\
U_{ij;kl} = \delta_{i+j,k+l}
\left(kC_{ij} - \frac{1}{6}(i+j)(k+1)k(k-1)(k-2)
+ \frac{1}{10}(k+2)(k+1)k(k-1)(k-2)\right),\ \ \ \
{\rm for}\ k\leq i,j, \nn \\
V_{ijk} = \frac{ijk}{3}\Big(i^2+j^2+k^2 +
3(ij+jk+ik) - 9(i+j+k) + 15\Big)
\ee
The complete expression for $U_{ij;kl}$ is restored by the symmetry
under the permutation $(i,j)\leftrightarrow (k,l)$. Note that the naive
continuation of formula (\ref{Us}) to the whole region of indices leads
to the non-symmetric $U_{ij;kl}$. Therefore, in this case
already $U_{ij;kl} = U_{ij;kl}^{(D)}$ is not
a polynomial of indices $i,j,k,l$.
Terms of the order $h^4$ in $A_\Pi$ still need to be calculated,
presumably, they will also be made from the same coefficients
$U_{ij;kl}$ and $V_{ijk}$ but with {\it a few} extra overall
coefficients as it happens to the $h^2$ and $h^3$ terms.

At least, the $\bar h$-linear terms in $D_\Pi$
with all possible
powers of $h$ can be summed up to give
\be
\oint(\bar z-\bar \zeta){\cal S}_\zeta\{z\}\zeta^2 d\zeta =
\oint \bar h(\bar \zeta)\left[\frac{h'''(\zeta)}{1+h'(\zeta)}
- \frac{3}{2}\left(\frac{h''(\zeta)}{1+h'(\zeta)}\right)^2\right]
\zeta^2 d\zeta
= \nn \\
= \sum_{p=1}^\infty (-)^{p-1}\!\!\!\!\!
\sum_{i_1,\ldots,i_p=0}^\infty
\frac{i_1\ldots i_p}{6p}
\left(\sum_{a=1}^p i_a^2 + 3\sum_{a<b}^p i_ai_b -
3p\sum_{a=1}^p i_a + \frac{p(3p+1)}{2}\right)
h_{1_1}\ldots h_{i_p}\bar h_{i_1+\ldots + i_p+1-p}
\ee
The above coefficients $A,C,V$ arise in particular
terms of this formula, with $p=1,2,3$ respectively and
\be
{\cal S}_\zeta\{z\} = \frac{z'''}{z'} - \frac{3}{2}
\left(\frac{z''}{z'}\right)^2 =
-\frac{1}{{z'}^2}{\cal S}_z\{\zeta\}, \ \ \ \ \ \
{\cal S}_\zeta\{z\} \frac{d\zeta}{\sqrt{dz/d\zeta}}
= -{\cal S}_z\{\zeta\}\frac{dz}{\sqrt{d\zeta/dz}}
\ee
is the Schwarzian derivative, which vanishes
identically for rational transformations
$z = \zeta+h(\zeta) = \frac{a\zeta+b}{c\zeta+d}$.
Of course, there is a complex conjugate contribution
which is linear in $h$ and sums up all possible powers
of $\bar h$.
It is unclear if a similar {\it local} expression can be
found for all other terms $h^p\bar h^q$ in $D_\Pi$
with both $p,q\geq 2$.
Even less clear is the situation with $A_\Pi$.

We use $\lambda$ and $\mu$ regularizations at the l.h.s. and at the
r.h.s. of (\ref{sgd}) respectively: \be D_\Pi = \left(\oint_{\phantom{a}\Pi}\oint
\frac{d\vec y d\vec y'}{(\vec y-\vec y')^2 + \lambda^2}\right) \
\stackrel{\lambda = \mu}{=}\  \int \frac{\sqrt{|\partial
H|^2(|\partial H|^2+|\partial r|^2)}} {r^2+\mu^2}d^2\zeta = A_\Pi
\label{lmreg} \ee Divergent contributions are proportional to the
length of the curve $\Pi$, \be\label{lenc} \frac{L}{2\pi} = 1 +
\frac{1}{2}(h_1+\bar h_1) - \frac{1}{8}(h_1^2+\bar h_1^2) +
\frac{1}{4}\sum_{k=1}^\infty {k^2|h_k|^2\over \left|1+h_1\right|^2}
- \nn \\-{1\over 4}h_1\bar h_1\left(h_1+\bar h_1\right) +
\frac{1}{16}(h_1^3+\bar h_1^3) - \frac{1}{16}\sum_{k,l=2}^\infty
kl(k+l-1) \left[{h_kh_l\bar h_{k+l-1}\over
\left(1+h_1\right)^2\left(1+\bar h_1\right)} + {\bar h_k\bar h_l
h_{k+l-1}\over \left(1+\bar h_1\right)^2\left(1+ h_1\right)}\right]
+ O(h^4) = \nn \\
= \sqrt{1+h_1}\sqrt{1+\bar h_1}\left(1 + \frac{1}{4}\sum_{k=2} {k^2
|h_k|^2\over \left|1+h_1\right|^2} - \frac{1}{16}\sum_{k,l=2}^\infty
kl(k+l-1) \left[{h_kh_l\bar h_{k+l-1}\over
\left(1+h_1\right)\left|1+h_1\right|^2} + {\bar h_k\bar h_l
h_{k+l-1}\over \left(1+\bar h_1\right)\left|1+h_1\right|^2}\right]+
\ldots\right) \ee

To summarize, the functional dependencies on {\it arbitrary} shape
of the curve $\Pi$ in $D_\Pi$ and $A_\Pi$ are almost the same, but
some overall coefficients are different, moreover, the number of
different coefficients can grow with the order of $h$-corrections.

It is unclear if this difference can be somehow
absorbed into the change of regularization
prescriptions.
Moreover, if instead of $\mu$-regularization at
the r.h.s. of (\ref{lmreg}), one cuts the area
integral at $|\zeta| = 1-c$, the answer for $A_\Pi$
changes drastically, leaving no observable similarity
to $D_\Pi$.
Worse than that, while the IR-finite part of $D_\Pi$
is invariant w.r.t. the
projective transformations $\delta z = \epsilon_- +
\epsilon_0 z + \epsilon _+ z^2$, i.e. is annihilated
by the three $SL(2)$ generators
\be
\hat J_{-}  = \frac{\partial}{\partial h_0}, \nn \\
\hat J_0 = \frac{\partial}{\partial h_1} + \sum_{k=0}^\infty
h_k\frac{\partial}{\partial h_k}, \nn \\
\hat J_{+} = \frac{\partial}{\partial h_2} + 2\sum_{k=0}^\infty h_k
\frac{\partial}{\partial h_{k+1}} + \sum_{k,l=0}^\infty
h_kh_l\frac{\partial}{\partial h_{k+l}}, \label{sl2gens} \ee this is
not true for the IR-finite part of $A_\Pi$ (actually in the $h^3$
approximation $\hat J_+ A^{finite}_\Pi\neq 0$ only because of a
wrong coefficient in front of a single term $h_2^2\bar h_3$, but
there can be more such bad terms when the power of $h$ increases).
In fact, this does not immediately contradict the conformal
invariance of $A_\Pi$, proved in \cite{koma}: the conformal symmetry
of \cite{koma} acts on $A_\Pi$ in a more sophisticated way than
(\ref{sl2gens}).

We refrain from making far-going conclusions from these
surprising results before they are independently checked.
In case if they are confirmed, they need and can be
straightforwardly extended in two obvious directions:
to higher orders in $h$-expansion and to "wavy lines".
This can help to better understand the structure of the
difference between $A_\Pi$ and $D_\Pi$ and hopefully find
a simple formulation of {\bf the anomaly in the Alday-Maldacena
duality} (\ref{sgd}).
Of course, this anomaly should be also extended to finite-$n$
polygons $\Pi$.
We emphasize that the apparent similarity between
(\ref{Aans}) and (\ref{Dans}) does not allow one to simply
reject (\ref{sgd}) (say, by claiming the failure of
the BDS conjecture), the relation looks too close to truth
to be simply ignored: one should rather search for
overlooked {\it corrections}, which can be responsible for
the {\it small} discrepancy between the l.h.s. and the r.h.s.
of (\ref{sgd}).

\subsection{Plan of the paper}

Below in this paper we provide a rather detailed derivation of
formulas (\ref{Dans}) in s.\ref{dint} and (\ref{Aans}) in
s.\ref{minar}, ending up with two simple MAPLE programs which can be
used for double-check and generalizations. These derivations are
preceded in s.\ref{plan} by a plan of such calculation, commenting
on various semi-technical issues, which can be useful for further
generalizations. Then, there is a brief discussion of global
conformal symmetry in s.5. Finally, the four Appendices contain the
derivation of formula for the circumference of the wavy circle and
other local counterterms in terms of parameters of the conformal
map, an alternative calculation of $D_{\Pi}$ using a different
regularization, a discussion of another, rectangular example that
allows one to test formula (\ref{sgd}), \cite{am3} and two MAPLE
programs that allow one to calculate $A_\Pi$ and $D_\Pi$.

\section{Wavy circle: the scheme of calculations
\label{plan}}
\setcounter{equation}{0}

\subsection{NG equation for $y_0=0$}

NG action with $y_0=y_3=0$ is quite simple,
\be
\int \frac{\sqrt{1 + (\partial_1 r)^2 + (\partial_2 r)^2}}
{r^2}\,dy_1 dy_2
\label{NGac1}
\ee
and equation of motion is:
\be
r\partial^2 r + 2(\partial r)^2 + 2 +
r\partial_i r\partial_j r \Big(\delta_{ij} \partial^2 r -
\partial^2_{ij} r\Big) = 0
\label{NGeqninf1}
\ee
or
\be
r\partial_i r\partial_j r \partial^2_{ij} r
= \Big(1 + (\partial r)^2\Big)\Big(2+r\partial^2r\Big)
\ee

There are a few exactly solvable examples that satisfy both the NG
equation (\ref{NGeqninf1}) and the boundary condition $r(y^2=1)=0$.
Unfortunately, they do not possess free parameters that can
be used to actually compare the l.h.s. and the r.h.s. of
(\ref{sgd}). In particular, the surface
\be\label{NGcirc}
r^2 = R^2-y^2 =R^2-y_1^2-y_2^2 = R^2-z\bar z
\ee
is a solution to  (\ref{NGeqninf1}).
It, indeed, provides a minimum of the regularized action. Later on, we
put $R=1$. In fact, changing $R$ is the zero-mode generated by the
coefficient $h_1$ of the conformal map. In fact, as illustrated by
(\ref{lenc}), $h_1$ enters all formulae in a special way, different from all
other $h_k$. Therefore, for the sake of
simplicity, we always put $h_1=\bar h_1=0$ and restore non-vanishing $h_1$ and
$\bar h_1$ only in s.5.

\subsection{Wavy circle: area calculation}

Here we consider an arbitrary infinitesimally deformed circle and
describe how to calculate its regularized minimal area in the
first non-trivial -- quadratic -- order in deformation
parameters.

To this end, we need to resolve the following problems:

\begin{itemize}

\item Choose an adequate parametrization of the deformation.
We do this by considering the conformal map $z = H(\zeta)$
of interior of the unit circle in the complex $\zeta$-plane
into the domain bounded by the deformed curve $\Pi$ in the complex
$z$-plane.
The map is an infinitesimal deformation of the unit map,
$H(\zeta) = \zeta + h(\zeta)$ and
\be
h(\zeta) = \sum_{k=0}^\infty h_k\zeta^k
\label{hpar}
\ee
is a small
function-valued parameter.

\item Find the shape of the minimal surface
$r^2(z,\bar z) = 1-\zeta\bar\zeta + a(\zeta,\bar\zeta)$
by solving the NG equation for $a(\zeta,\bar\zeta)$ and imposing
the boundary condition
\be
a\Big(e^{i\phi},e^{-i\phi}\Big)=0
\label{boco}
\ee
For $h\neq 0$ vanishing everywhere $a=0$ is not a solution, and
one needs to calculate $a$ up to the second order in $h$.
The relevant form of the NG equation in this case is
\be
\Delta_{NG}\Big(a + u(h)\Big) = O(a^2,ah,h^2)
\label{NGlin}
\ee
where $\Delta_{NG}$ is a linear differential operator
(already found in \cite{im8})
\be
\Delta_{NG}=
\Delta_0 - {\cal D}^2 + {\cal D} =
4\partial\bar\partial - \bar z^2\bar\partial^2 -
2z\bar z \partial\bar\partial - z^2\partial^2
\ee
expressed through the ordinary Laplace and dilatation
operators $\Delta_0 = \partial^2_1+\partial_2^2$ and
${\cal D} = y_1\partial_1 + y_2\partial_2$, and
\be
u(h) = 2\zeta\bar\zeta
\sum_{k=1}^\infty {\rm Re}\Big(h_k\zeta^{k-1}\Big)
\label{uthrh}
\ee
is linear in $h$.

\begin{itemize}

\item[{\bf A.}]
As a first step towards solving (\ref{NGlin})
we can put $h=0$ and neglect the quadratic term ${\cal Q}(a)$,
i.e. consider the equation
\be
\Delta_{NG}(a)=0
\label{NG0}
\ee
Its generic solution was found in \cite{imm,im8} in the form
\be
a(\zeta,\bar\zeta) = 2\sum_{k=0}^\infty {\rm Re}\Big(a_k\zeta^k\Big)
F_k(\zeta\bar\zeta)
\label{athrF}
\ee
where
\be
F_k(x) = \frac{(1+k\sqrt{1-x})(1-\sqrt{1-x})^k}{x^k}
\sim
\phantom._2F_1\left(\frac{k}{2},\frac{k-1}{2};k+1;x\right)
\label{Fkhyp}
\ee
are specific hypergeometric functions expressed through the Legendre (spherical)
functions $Q^{-3/2}_{k-1/2}$
\be
\phantom._2F_1\left(\frac{k}{2},\frac{k-1}{2};k+1;x\right)=
2^kk(k-1)i\sqrt{{2}\over{\pi}}\left({1-x\over x}\right)^{3\over 4}
x^{1-k\over 2}Q^{-3/2}_{k-1/2}\left({1\over\sqrt{x}}\right)
\ee
We normalize $F_k(x)$
by the condition
\be
F_k(1) = 1,
\ee
i.e. divide the hypergeometric series at the r.h.s. of (\ref{Fkhyp})
by their values at $x=1$,
\be
\phantom._2F_1(a,b;c;1) = {\Gamma (c)\Gamma (c-a-b)\over
\Gamma (c-a)\Gamma (c-b)},\ \ \ \ \hbox{Re } c>\hbox{Re } b>0,\ \hbox{Re }
(c-a-b)>0 \\
\phantom._2F_1\left(\frac{k}{2},\frac{k-1}{2};k+1;1\right) = {2^k\over k+1}
\ee
In particular,
\be
F_0(x) = 1, \nn \\
F_1(x) = 1, \nn \\
F_2(x) = \frac{-2+3x+2(1-x)^{3/2}}{x^2}, \nn \\
F_3(x) = \frac{-8+12x-3x^2+8(1-x)^{3/2}}{x^3}, \nn \\
F_4(x) = \frac{-24+40x-15x^2+8(6-x)(1-x)^{3/2}}{x^4}, \nn \\
\ldots
\ee
In the vicinity of $x=1$ these $F_k$ behave as follows:
\be
{\rm at}\ \ \ x=1-c^2\ \ \ \ \ \ \ \ \ \ \ \
F_k= 1-\frac{k(k-1)}{2}c^2 - \frac{k(k^2-1)}{3}c^3
+ O(c^4)
\label{Fasy}
\ee

Of course, for $h=0$ the boundary condition (\ref{boco})
implies that in (\ref{athrF}) all $a_k=0$.

\item[{\bf B.}]
Since in neglect of its r.h.s. (\ref{NGlin}) differs
from (\ref{NG0}) only by a shift of $a$, we can
use the same result (\ref{athrF}) for $a+u(h)$.
Moreover, the explicit form  (\ref{uthrh}) of the shift $u(h)$
is very simple, so that one can easily impose the
boundary conditions (\ref{boco})
\be
a(\zeta,\bar\zeta) =
2\sum_{k=1}^\infty {\rm Re}\Big(h_k\zeta^{k-1}\Big)
A_k(\zeta\bar\zeta) + O(h^2), \nn \\
A_k(x) = F_{k-1}(x) - x
\label{athrh}
\ee
and, according to (\ref{Fasy}),
\be
{\rm at}\ \ \ x=1-c^2\ \ \ \ \ \ \ \ \ \ \ \
A_k = -\frac{k(k-3)}{2}\,c^2 + \frac{k(k-1)(k-2)}{3}\,c^3
+ O(c^4), \nn \\
A_k' = \frac{k(k-3)}{2} - \frac{k(k-1)(k-2)}{2}\,c
+ O(c^2)\
\label{Aasy}
\ee

\end{itemize}

\item Evaluate (regularized)
effective action up to the $h^2$-terms.
It diverges and we regularize it. It can be done in many different ways,
here we use the two most naive possibilities which are, however, representative
enough to illustrate the typical features. As we shall see, the result
drastically depends on the choice of regularization.

According to \cite{am1}, the regularization procedure implies modifying
the action but using the old solution (which is, definitely, a somewhat
controversial prescription).

According to \cite{AdSRG} the most appropriate way to regularize
AdS quantities is to make a shift away from the boundary at $r=0$ to
$r=\epsilon$:
dependence of the bulk action on the shift is the counterpart of
renormalization group for the boundary theory.

The question in our case is where we impose the vanishing
boundary conditions: on the boundary or on the shifted boundary?

Another question is what kind of shift we should perform:
it can be of an arbitrary shape and the corresponding renormalization
group is in fact infinite-dimensional \cite{mmRG}.
The conventional one-parametric renormalization subgroup corresponds
to a kind of a "constant" shift.

Of this large variety of possibilities,
we consider two different regularizations:

\paragraph{$c$-regularization:} boundary condition at the {\it original}
boundary, the shift is "constant", compare with RG of \cite{AdSRG}
and with \cite{am1}.
Implies drastic violation of (\ref{sgd}) in the case of deformed
circle. We make this regularization by cutting the integral
over $x\equiv\zeta\bar\zeta$
at $1-c^2$ with non-vanishing $c$,\ \footnote{Following \cite{am1},
one would also have to
introduce a $c$-dependent factor $\beta(c)=1+\beta_1c+O(c^2)$
into the integrand of action:
\be
\frac{
\sqrt{|\partial H|^2\Big(|\partial H|^2 + 4|\partial r|^2}\Big)
}{r^2} \to
\frac{\sqrt{|\partial H|^2
\Big(r^2|\partial H|^2 + \beta|\partial r^2|^2}\Big)
}{r^3}
\label{regac}
\ee
This, however, does not lead to any essential effects later on, and we
ignore such a modification here. In fact, the role of $\beta_1$ would
be just to shift
$\sigma^{reg}_j \rightarrow \sigma^{reg}_j + \frac{\beta_1}{2}
\sigma^{sing}_j$ in the formulas below.
In fact, $\beta_1$ has dimension $length^{-1}$ and can hardly be
constant.}

\be
S_{NG}\{a,h\} = \int_{|\zeta|^2\leq 1-c^2} \frac{
\sqrt{|\partial H|^2\Big(|\partial H|^2 + 4|\partial r|^2}\Big)
}{r^2}\,d^2\zeta =
\int_{|\zeta|^2\leq 1-c^2} \frac{
\sqrt{|\partial H|^2
\Big(r^2|\partial H|^2 + |\partial r^2|^2}\Big)
}{r^3}\,d^2\zeta
\label{regac1}
\ee

For
\be
r^2 = 1-|\zeta|^2+a(\zeta,\bar\zeta)
\ee
the action can be expanded as
\be
S_{NG}\{a,h\} =
\int_{|\zeta|^2\leq 1-c^2} \sqrt{\frac{|\partial H|^2
\Big(|\partial H|^2\big(1-|\zeta|^2\big) +
\beta|\zeta|^2 + |\partial H|^2a - \beta{\cal D}a
+ \beta|\partial a|^2\Big)
}{(1-|\zeta|^2+a)^{3/2}}}d^2\zeta = \nn \\
=S_{circ} + S_0\{h\} + S_1\{a,h\} + S_2\{a\} + O(a^{3-j}h^j)
\label{NGthra}
\ee
Here $S_j\{a,h\}$ is of degree $j$ in $a$ and of
degree $0,\ldots,2-j$ in $h$ and we specially distinguish the contribution
that does not depend on $h$ at all, $S_{circ}$.
As a function of regularization parameter $c$,
each
\be
S_j = \frac{1}{c}S_j^{sing} + S_j^{reg} + O(c)
\ee
After substitution of (\ref{athrh}) each $S_j$ becomes
a function of the boundary shape $h(z)$:
\be
S_j\{a(h),h\} =
2\pi\sum_{k=2}^\infty |h_k|^2 \sigma_k^{(j)} + O(h^3)
\ee

Terms of the order $O(c)$ are omitted,
with this accuracy one has
\be\label{sigmas}
\sigma^{(0)}_k = k^2\int_0^{1-c^2}\frac{x^{k-1}dx}{(1-x)^{3/2}}
\left(1-\frac{x}{2}-\frac{x^2}{4}\right) = \frac{k^2}{2c}+I_1,\nn\\
\sigma^{(1)}_k =\underbrace{ -\left.\frac{kA_k
x^{k+1}}{(1-x)^{3/2}}\right|_{x=1-c^2}} +
\frac{k}{2}\int_0^{1-c^2}\frac{A_kx^{k-1}dx}{(1-x)^{3/2}}
\Big((k+1)x-4\Big) = \underbrace{\frac{k^2(k-3)}{2c} -
\frac{k^2(k-1)(k-2)}{3}}
+ 2I_2,\nn\\
\sigma^{(2)}_k =
\underbrace{\left.\frac{x^{k-1}A_k^2}{2(1-x)^{5/2}}
\Big((k-2)x^2-2(k-3)x+(k-1)\Big)
\right|_{x=1-c^2}}
-  \frac{k}{4}\int_0^{1-c^2}\frac{A_kx^{k-1}dx}{(1-x)^{3/2}}
\Big((k+1)x-4\Big) = \nn \\
= \underbrace{\frac{3k^2(k-3)^2}{8c} - \frac{k^2(k-1)(k-2)(k-3)}{2}}
- I_2 \ee where we "underbraced" the boundary contributions (which come
from the integration by parts). The density integrals \be I_1 =
k^2\int_0^{1}\frac{dx}{(1-x)^{3/2}}
\left\{x^{k-1}\left(1-\frac{x}{2}-\frac{x^2}{4}\right) +
\frac{x-2}{4}\right\} = \frac{4k^2}{3}\left(1 +
3\sum_{j=1}^{k-1}(-)^jC^j_{k-1}
\frac{2j^2+3j-1}{(2j-1)(2j+1)(2j+3)}\right),
\ee
\be
I_2 = \frac{k}{2}\int_0^{1}\frac{dt}{t^2}
\Big(\big(1+(k-1)t\big)(1-t)^{k-1} - (1-t^2)^k\Big)
\Big((k-3) - (k+1)t^2\Big)
\ee
are rather complicated, however, their sum is simple:
\be
I_1 + I_2 = -\frac{k(k-1)(k-2)}{2}
\ee
Non-transcendental boundary terms contribute
\be
 - \frac{k^2(k-1)(k-2)}{3} - \frac{k^2(k-2)(k-2)(k-3)}{2} =
- k(3k-7)\frac{k(k-1)(k-2)}{6} \ee To summarize, the singular term
is \be\label{sigmasing} \sigma_k^{sing} =
\frac{k^2(3k^2-14k+19)}{8c} = \frac{k^2}{2c} +
\underbrace{\frac{3k^2(k^2-4k+5)}{8c} - \frac{k^3}{4c}}
\ee while the regular term is \be \sum_{k=2}^\infty
\sigma_k^{reg}|h_k|^2 = -\sum_{k=3}^\infty
(\underbrace{3k^2-7k}+3)\frac{k(k-1)(k-2)}{6}|h_k|^2 \ee

\paragraph{$\mu$-regularization:} the better one, no direct relation
to \cite{AdSRG}, provides (\ref{sgd}) for deformed circle up
to the coefficient $3$ in front of the $h^2$ terms. The regularization implies
just replacing $r^2\to r^2+\mu^2$ in the denominator of
the integrand of action:

\be
S_{NG}\{a,h\} = \int_{|\zeta|^2\leq 1} \frac{
\sqrt{|\partial H|^2\Big(|\partial H|^2 + 4|\partial r|^2}\Big)
}{r^2+\mu^2}\,d^2\zeta =
\int_{|\zeta|^2\leq 1} \frac{
\sqrt{|\partial H|^2
\Big(r^2|\partial H|^2 + |\partial r^2|^2}\Big)
}{r(r^2+\mu^2)}\,d^2\zeta
\label{regac2}
\ee

In the case of $\mu$-regularization the boundary (underlined) terms
in (\ref{sigmas}) do {\bf not} contribute, and the full answer is
\be {\rm Area}_\mu = \frac{\pi^2}{\mu}
\left(1+\frac{1}{4}\sum_{k=2}^\infty k^2|h_k|^2\right) +
\pi\left(I_1+I_2\right) - 2\pi + O(h^3) \ee The combination
$\left({\pi^2\over\mu} -2\pi\right)$ here is $S_{circ}$.

\bigskip

One would expect that the divergent part of the result should be
proportional to the length of the wavy circle. This is, indeed, the
case for the $\mu$-regularization, since the length of the contour
is (see Appendix I) \be {L\over 2\pi} = \oint_\Pi dl = 1 +
\sum_{k=2}^\infty \frac{k^2|h_k|^2}{4} + O(h^3) \ee and \be {\rm
Area}_\mu =\frac{\pi L}{2\mu}- \pi\sum_{k=3}^\infty
\frac{k(k-1)(k-2)}{2}|h_k|^2 -2\pi+ O(h^3) \ee

\paragraph{Comment.} Note that
the result for the $c$-regularization is not same good. It is not
proportional to the length and, what is much worse, one can hardly
find local boundary counterterms to treat the singularity. Indeed,
the only other possible candidate could be integral of logarithm of
the scalar curvature \be \kappa = \frac{\left|{\rm Im}(\ddot
z\dot{\bar z})\right|}{|\dot z|^3} \ee However, this integral \be
\oint_\Pi \log \kappa dl \sim \sum_{k=2}^\infty k^2(k^2-4k+5)|h_k|^2
\ee along with the length term, leave the unbalanced singular term
$-\sum {k^3\over 4c}|h_k|^2$, see (\ref{sigmasing}).

\end{itemize}

\paragraph{The main lesson}one can get from this consideration is that,
generally speaking, the result strongly depends on the regularization
procedure. However, we expect that for the class of admissible regularizations,
i.e. such that the surface terms vanish, the result for the finite part of
the minimal area would not depend on the regularization.
For example, our $c$-regularization implied that the
boundary condition is set at the original boundary. One can instead shift
the boundary conditions to the regularized boundary what effectively
corresponds to omitting
the surface terms. As we saw above, this would lead to the same result as
for the $\mu$-regularization.

\subsection{Double countour integral}

Above results for the area should now be compared with the
(regularized) double loop integral evaluated with the same accuracy
up to the $h^2$-terms.
The result for the finite piece is\footnote{To obtain this result, one uses
the following integrals:
\be
\int\left(\frac{\sin (k\varphi)}{\sin\varphi}\right)^2\,
\frac{d\varphi}{2\pi} = k, \ \ \ \ \
\int\frac{\sin\big((2k+1)\varphi\big)-(2k+1)\sin\varphi}
{\sin^3\varphi}\, \frac{d\varphi}{2\pi} = -2k(k+1)\\
\int \frac{\cos\big((k+1)\varphi\big)\sin (k\varphi) -
k\sin\varphi}{\sin^3\varphi}\,\frac{d\varphi}{2\pi} = -k(k+1),\ \ \ \ \ \
\int \frac{\cos(2\varphi)\sin^2(k\varphi) - k^2\sin^2\varphi}
{\sin^4\varphi}\, \frac{d\varphi}{2\pi} = -\frac{2k(k^2+2)}{3}
\ee
}

\be \sum_{k=2}^\infty |kh_k|^2 \int
\frac{d\varphi}{4\sin^2\!\varphi} \Big(\cos(2k\varphi) -
2\sigma_k(\varphi)\cos\big((k+1)\varphi\big)
+ \sigma^2_k(\varphi)\cos(2\varphi)\Big) 
= -2\pi\sum_{k=3}^\infty \frac{k(k-1)(k-2)}{6}\,|h_k|^2 \ee Here
$\sigma_k(\varphi) \equiv \frac{\sin k\varphi}{k\sin\varphi}$. The
divergent piece (see details in (\ref{dp}) below) is
$\frac{L}{\lambda}$, for example, from \be \int_0^{2\pi}\frac{B
d\varphi}{B\sin^2\varphi + \lambda^2} =
\frac{2\pi\sqrt{B}}{\lambda}+O(\lambda) \ee and $\oint
\sqrt{B(\Phi)}d\Phi = L$. No term $\ \ \lambda^{-1}\oint\log
\kappa\, dl\ \ $  is present.

\subsection{Double integral vs. minimal area}

Now, comparing the results of two calculations for the double contour
integral and for the area, one can see another problem with the
$c$-regularization: the finite piece it gives
has nothing to do with the result for the double contour
integral (\ref{Dans}).  Indeed,

\be
\sigma^{reg}_k - D_k =
-(3k^2-7k+4)\frac{k(k-1)(k-2)}{6} + \frac{k(k-1)(k-2)}{6}
=-(3k^2-7k+2)\frac{k(k-1)(k-2)}{6} = \nn \\
= \frac{k(3k-1)(k-1)(k-2)^2}{6}
\ee
which can not be removed into ${\beta_1\over 2}\sigma^{sing}_k$.

\bigskip

At the same time, the case of ${\rm Area}_\mu$ is much better,
though differs by a factor from the double integral: \be\label{A2}
{\rm Area}_\mu = \frac{\pi L}{2\mu} - \pi\sum_{k=3}^\infty
\frac{k(k-1)(k-2)}{2}|h_k|^2-2\pi + O(h^3) \ee while \be\label{D2}
{\rm D}_\lambda =  \frac{2\pi L}{\lambda} -2(2\pi)^2
\sum_{k=3}^\infty \frac{k(k-1)(k-2)}{6}|h_k|^2 -(2\pi)^2+ O(h^3) \ee

By all these reasons, we choose in further calculations only the
$\mu$-regularization, keeping in mind that the final result can
drastically depend on the regularization, and it is not guaranteed
that the $\mu$-regularization is the best/correct one. Anyhow, in
this regularization a discrepancy in the overall coefficient occurs
in the $h^2$ terms between (\ref{A2}) and (\ref{D2}). If our
argument at the end of s.2.2 about regularization independence of
${\rm Area}_\mu$ is taken seriously, this discrepancy is unavoidable
and becomes a kind of anomaly, slightly violating the conjectured
form (\ref{sgd}) of the Alday-Maldacena duality.

\subsection{Further corrections}

The next step is to check if the same discrepancy is presented in higher orders
in $h$.
Naively,
one would expect that, in order to obtain $h^3$-corrections to $A_\mu$, one
needs to take into account higher terms in the NG equation etc.
However, it turns out that these corrections can be obtained with
the already obtained solution (\ref{athrF}). To see this, let us introduce the
notation $S_{(k,l)}$ for the term in action of the order $a^kh^l$. Then, up to
the third order, the action is
\be
S=\sum_{l,k=1}^3S_{(l,k)}
\ee
and the solution to the equation of motion $a^{(1)}$
linear in $h$ is determined from the variation (note that $S_{(1,0)}=0$)
\be\label{eqm}
\left.\left({\delta S_{(1,1)}\over\delta a}+{\delta S_{(2,0)}\over\delta a}
\right)\right|_{a=a^{(1)}}=0
\ee
In order to find the next correction, $a^{(2)}$ one needs to insert
$a=a^{(1)}+a^{(2)}$ into the equation
\be
{\delta S_{(1,1)}\over\delta a}+{\delta S_{(2,0)}\over\delta a}+
{\delta S_{(2,1)}\over\delta a}+{\delta S_{(1,2)}\over\delta a}
+{\delta S_{(3,0)}\over\delta a}=0
\ee
etc. Now one needs to calculate the
value of action (i.e. the minimal area) on the solution
\be
A=\sum_{l,k=1}^3S_{(l,k)}\left(a^{(1)}+a^{(2)}+\ldots\right)=A^{(2)}+A^{(3)}
+\ldots
\ee
Note that the part of the cubic correction $A^{(3)}$ that involves $a^{(2)}$
is linear in it, and, therefore, is proportional to
$\left.\left({\delta S_{(1,1)}\over\delta a}+
{\delta S_{(2,0)}\over\delta a}\right)
\right|_{a=a^{(1)}}$ which vanishes by the equation
of motion, (\ref{eqm}). Therefore, only $a^{(1)}$ contributes to the minimal
area up to the third order, and one can use the known solution,
(\ref{athrF}) when evaluating the minimal area.

Thus, one just needs to insert solution (\ref{athrF}) into the
action and expand it up to $h^3$ terms. Similarly, one needs to
calculate the double contour integral $D_\mu$ up to terms of the
same cubic order. This can be done by pen, or with the computer (the
corresponding MAPLE programs can be found in Appendix IV), the
results being formulas (\ref{Aans}) and (\ref{Dans}). In the latter
case, the $h^2$ terms and some other contributions of higher order
are also presented in order to give a flavour of how they look like.
However, in order to include higher order (quartic) terms into the
expression for $A_\mu$, one would need to find corrections to the NG
solution which is a tedious problem. Here we restrict ourselves only
to the cubic terms.

\section{Double integral: technicalities
\label{dint}}
\setcounter{equation}{0}

\subsection{BDS formula and double loop integral}

In the (homogeneous) $n=\infty$ case the BDS formula immediately
leads to the double integral, hence, the calculation of \cite{bht}
can be bypassed.
According to \cite{mmt1}, the BDS formula is a sum over $4$-boxes
and each $4$-box degenerates into a chordae of the curve $\Pi$
when $n\rightarrow \infty$.
The contributions of each $4$-box consists of dilogarithmic and
logarithmic parts, which degenerate into
\be
Li_2 \Big(1 - \exp (\tau_l+\tau_s-\tau_{m1}-\tau_{m2})\Big)
\stackrel{n\rightarrow\infty}{\longrightarrow}
Li_2 \left(1 - \exp \left(\delta\phi\delta\phi'
\frac{\partial^2 \log t(\phi,\phi')}{\partial \phi\partial\phi'}
\right)\right) = \nn \\
= \delta\phi\delta\phi'
\frac{\partial^2 \log t(\phi,\phi')}{\partial \phi\partial\phi'}
+ O(\delta\phi^3)
\ee
and
\be
(\tau_l-\tau_{m1})(\tau_l-\tau_{m2})
\stackrel{n\rightarrow\infty}{\longrightarrow}
\delta\phi\delta\phi'
\frac{\partial\log t(\phi,\phi')}{\partial\phi}
\frac{\partial\log t(\phi,\phi')}{\partial\phi'}
+ O(\delta\phi^3)
\ee
respectively. Here $t = |z(\phi)-z(\phi')|^2$ is the squared
length of the chordae, $\tau = \log t$.
Adding the dilogarithmic and logarithmic contributions and summing over
chordae, one straightforwardly reproduces the double contour integral
\be
\oint\oint_\Pi d\phi d\phi' \left(
\frac{\partial^2 \log t(\phi,\phi')}{\partial \phi\partial\phi'}+
\frac{\partial\log t(\phi,\phi')}{\partial\phi}
\frac{\partial\log t(\phi,\phi')}{\partial\phi'}\right)
= \oint\oint_\Pi d\phi d\phi' \frac{1}{t}
\frac{\partial^2 t(\phi,\phi')}{\partial \phi\partial\phi'}
= \oint\oint_\Pi \frac{{\rm Re}(d\vec y d\vec y')}{(\vec y-\vec y')^2}
\ee

\subsection{Double loop integral for the wavy circle}

The necessary ingredients of the double integrand are \be |z-z'|^2 =
4\sin^2\varphi \Big\{1 + \sum_k k \left(h_ke^{i(k-1)\Phi} + \bar
h_ke^{-i(k-1)\Phi}\right) \sigma_k(\varphi) + \sum_{k,l} klh_k\bar
h_l \sigma_k(\varphi)\sigma_l(\varphi) e^{i(k-l)\Phi}\Big\} \ee with
$\phi=\Phi-\varphi$, $\phi'=\Phi+\varphi$, $\sigma_k(\varphi) =
\frac{\sin k\varphi}{k\sin\varphi}$ and
$$
\frac{1}{2}(dzd\bar z' + d\bar zd z') =
$$
\vspace{-0.5cm} \be = 2d\Phi d\varphi \left\{  \cos(2\varphi) +
\sum_k k \left(h_ke^{i(k-1)\Phi}+\bar h_ke^{-i(k-1)\Phi}\right)
\cos(k+1)\varphi +\sum_{k,l} klh_k\bar h_l
e^{i(k-l)\Phi}\cos(k+l)\varphi \right\} \ee Now one needs to
regularize the integral and, then, to calculate it (we remind that
$h_1=\bar h_1=0$ to simplify formulae) \be\label{DIc}
D_{\Pi}={1\over 2}\oint\oint \frac{dzd\bar z' + d\bar zd
z'}{|z-z'|^2+\lambda^2}=
\\
=2\oint d\Phi \oint d\varphi \frac{\cos(2\varphi) + \sum_k k
\left(h_ke^{i(k-1)\Phi}+\bar h_ke^{-i(k-1)\Phi}\right)
\cos(k+1)\varphi +\sum_{k,l} klh_k\bar h_l
e^{i(k-l)\Phi}\cos(k+l)\varphi} {4\sin^2\varphi\left(1 + \sum_k k
\left(h_ke^{i(k-1)\Phi} + \bar h_ke^{-i(k-1)\Phi}\right)
\sigma_k(\varphi) + \sum_{k,l} klh_k\bar h_l
\sigma_k(\varphi)\sigma_l(\varphi)
e^{i(k-l)\Phi}\right)+\lambda^2}=\\= 2\oint d\Phi \oint
\frac{d\varphi \cos(2\varphi)}{4\sin^2\varphi +{\lambda^2\over
B(\varphi,\Phi)}} +4\pi\sum_{k=1}^\infty |kh_k|^2 \int
\frac{d\varphi}{4\sin^2\!\varphi} \Big(\cos(2k\varphi) -
2\sigma_k(\varphi)\cos\big((k+1)\varphi\big) +
\sigma^2_k(\varphi)\cos(2\varphi)\Big)\\\vspace{-0.3cm}
B(\varphi,\Phi)\equiv 1 + \sum_k k \left(h_ke^{i(k-1)\Phi} + \bar
h_ke^{-i(k-1)\Phi}\right) \sigma_k(\varphi) + \sum_{k,l} klh_k\bar
h_l \sigma_k(\varphi)\sigma_l(\varphi) e^{i(k-l)\Phi}\label{ssf} \ee
\vspace{-0.3cm} The second term in (\ref{ssf}) is finite, we
discussed it above in ss.2.3, while the first one diverges and is
equal to \vspace{0.3cm} \be\label{dp} 2\oint d\Phi \oint
\frac{d\varphi \cos(2\varphi)}{4\sin^2\varphi +{\lambda^2\over
B(\varphi,\Phi)}}=2\oint d\Phi \left({\pi\sqrt{B(0,\Phi)}
\over\lambda}-\pi\right)+O(\lambda)=4\pi\left({\pi\over\lambda}\left[
1+{1\over 4}\sum_k k^2|h_k|^2\right]-\pi\right)+O(\lambda)=\\
={2\pi L\over\lambda}-4\pi^2+O(\lambda) \ee The constant $4\pi^2$
can be removed, e.g., by the proper choice of $\beta_1$ (see
footnote 2) and we ignore it from now on.

One can also try other regularizations in calculating the double
loop integral. However, as we demonstrate in Appendix II, using a
counterpart of the $c$-regularization does not change the result.

\section{Minimal area: technicalities
\label{minar}}
\setcounter{equation}{0}

Here we reproduce some technicalities of calculation of the minimal area
skipped in section 2.

First of all, we construct the solution to the NG equation in the
second order in $h$ and, then, expand the action up to the same second order
and reduce the integrals emerging to (\ref{sigmas}).

\subsection{Approximate NG equation}

We are interested in the contribution $\sim h^2$ to the
regularized NG action.
Solving the NG equation we obtain
\be
a = (1-\zeta\bar\zeta)\left(1 + \sum_{k\geq 0}{\rm Re}(a_kh_k)
+ \sum_{k,l\geq 0}{\rm Re}(a_{kl}h_kh_l + \tilde a_{kl}h_k\bar h_l)
+ O(h^3)\right)
\ee

For $\partial H=1$:
\be
\Delta_{NG} a \equiv
\Big(\Delta_0 - {\cal D}^2 + {\cal D}\Big)a =
\Big(4\partial\bar\partial - \zeta^2\partial^2
-2\zeta\bar\zeta\partial\bar\partial -
\bar\zeta^2\bar\partial^2\Big)a =
O(a^2)
\ee
or, with $a^2$-terms included,
\be
\Delta_{NG} a
+ 2\Big( \zeta\bar\partial a \partial^2 a
+ \bar\zeta\partial a\bar\partial^2a -
(\zeta\partial a + \bar\zeta\bar\partial a)
\partial\bar\partial a\Big)
+ \frac{a}{1-\zeta\bar\zeta}\Big(
\zeta^2\partial^2 -2\zeta\bar\zeta\partial\bar\partial +
\bar\zeta^2\bar\partial^2\Big)a = O(a^3)
\ee
i.e.
\be
\Delta_{NG} a + {2}{\cal D}(\partial a\bar\partial a)
- ({\cal D}a)\Delta_0 a + a\Delta_0 a -\frac{1}{1-\zeta\bar\zeta}
a\Delta_{NG} a = O(a^3)
\label{H1aliqua}
\ee
(note that $\Delta_0 = \partial_1^2+\partial_2^2
= 4\partial\bar\partial$,
$\ (\partial_i a)^2 = 4\partial a\bar\partial a$,
$\ {\cal D} = \zeta\partial + \bar\zeta\bar\partial\ $
and $\ \zeta^2\partial^2 +2\zeta\bar\zeta\partial\bar\partial +
\bar\zeta^2\bar\partial^2 = {\cal D}^2 - {\cal D}$).

Now we switch on $\partial H\neq 1$:
$$
\frac{|\partial H|^2}{4(1-\zeta\bar\zeta)}\left\{
2(|\partial H|^2-1)\Big(\zeta\bar\zeta
+ 2|\partial H|^2(1-\zeta\bar\zeta)\Big) -
\zeta\bar\zeta{\cal D}\left(\log(|\partial H|^2)\right) + \right.
$$
\centerline{$
+ \Delta_{NG} a  +
(|\partial H|^2-1)(1-\zeta\bar\zeta)\Delta_0a
+\left({\cal D}a+\frac{\zeta\bar\zeta a}{1-\zeta\bar\zeta}
\right)\Big(2(1-|\partial H|^2)+
{\cal D}\left(\log(|\partial H|^2)\right)\!\Big)
+ \zeta\bar\zeta\Big(\partial a \bar\partial \log|\partial H|^2 +
\bar\partial a \partial \log|\partial H|^2\Big) +
$}
\be
\left. + {2}{\cal D}(\partial a\bar\partial a)
- ({\cal D}a)\Delta_0 a + a\Delta_0 a -\frac{1}{1-\zeta\bar\zeta}
a\Delta_{NG} a\right\} = O(a^kh^{3-k})
\label{NGeqah}
\ee
The $a$-independent piece in curved brackets in (\ref{NGeqah}) is
\be
8\sum_{k=1}^\infty {\rm Re} \left(kh_k \zeta^{k-1}\right)
\left(1-\frac{k+1}{4}\zeta\bar\zeta\right) + O(h^2) =
2\Delta_{NG} \left(\sum_{k=1}^\infty {\rm Re}
\Big(\bar\zeta h_k\zeta^k\Big)\right)
\ee
Thus (see (\ref{athrh}))
\be
a(\zeta,\bar\zeta) = 2\sum_{k=1}^\infty
{\rm Re}\Big(h_k\zeta^{k-1}\Big)
\left(\frac{F_{k-1}(\zeta\bar\zeta)}{F_{k-1}(1)}
- \zeta\bar\zeta\right)
+ O(h^2)
\label{avsh}
\ee
(this quantity vanishes when $\zeta\bar\zeta=1$ and
$ a(\zeta,\bar\zeta) + \bar\zeta h(\zeta) + \zeta \bar h(\bar\zeta)
= a(\zeta,\bar\zeta)
+ 2\sum_{k=1}^\infty {\rm Re} \big(\bar\zeta h_k\zeta^k\big)$
is a zero-mode of $\Delta_{NG}$).

\subsection{NG action on NG solution up to the $h^2$ terms}

\be
\int \frac{\sqrt{|\partial H|^2\big(|\partial H|^2
+ 4\partial r\bar\partial r\big)}}{r^2}\,d^2\zeta
= \int \frac{\sqrt{|\partial H|^2\Big(|\partial H|^2
(1-\zeta\bar\zeta + a) + \big|\zeta-\bar\partial a\big|^2\Big)}}
{r^3}\,d^2\zeta=
\\
=\int \frac{\sqrt{|\partial H|^2\Big(|\partial H|^2 +
\zeta\bar\zeta \big(1-|\partial H|^2\big) +
a|\partial H|^2 - {\cal D}a + |\partial a|^2
\Big)}}{(1-\zeta\bar\zeta)^{3/2}}
\left(1-\frac{3a}{2(1-\zeta\bar\zeta)}
+ \frac{15a^2}{8(1-\zeta\bar\zeta)^2} + O(a^3)\right)
d^2\zeta
\ee
At the moment we ignore regularization, it can be easily restored.
Under the root sign one has up to the second order
in $h$:
$$
\Big(1+(\partial h+\overline{\partial h}) + |\partial h|^2\Big)
\Big(1+(\partial h+\overline{\partial h})(1-\zeta\bar\zeta)
+ |\partial h|^2(1-\zeta\bar\zeta)  +  (a-{\cal D}a)
+ |\partial a|^2 + a(\partial h+\overline{\partial h}) \Big)
=$$
$$
= 1 + \Big(a-{\cal D}a
+ (\partial h+\overline{\partial h})(2-\zeta\bar\zeta)\Big)
+ \Big((\partial h+\overline{\partial h})^2(1-\zeta\bar\zeta)
+ |\partial h|^2(2-\zeta\bar\zeta)
+ 2(\partial h+\overline{\partial h})a
-(\partial h+\overline{\partial h}){\cal D}a)
+ |\partial a|^2\Big)
$$
and the square root is equal to
$$
1 + \frac{1}{2}\Big(a-{\cal D}a
+ (\partial h+\overline{\partial h})(2-\zeta\bar\zeta)\Big) +
$$
$$
+\frac{1}{8}\Big(
4(\partial h+\overline{\partial h})^2(1-\zeta\bar\zeta)
+ 4|\partial h|^2(2-\zeta\bar\zeta)
+ 8(\partial h+\overline{\partial h})a
-4(\partial h+\overline{\partial h}){\cal D}a)
+ 4|\partial a|^2
-\big(a-{\cal D}a
+ (\partial h+\overline{\partial h})(2-\zeta\bar\zeta)\big)^2
\Big) =
$$
$$
= 1 + \frac{1}{2}\Big(a-{\cal D}a
+ (\partial h+\overline{\partial h})(2-\zeta\bar\zeta)\Big) +
\frac{1}{8}\Big(4|\partial a|^2 - (a-{\cal D}a)^2\Big) +
$$
$$
+ \frac{1}{8}\Big(
4|\partial h|^2(2-\zeta\bar\zeta)
-(\zeta\bar\zeta)^2(\partial h+\overline{\partial h})^2
+2(2+\zeta\bar\zeta)(\partial h+\overline{\partial h})a
-2\zeta\bar\zeta(\partial h+\overline{\partial h}){\cal D}a
\Big)
$$
Now we substitute
\be
\partial h = \sum_{k=1}^\infty kh_k\zeta^{k-1}, \ \ \ \ \
\partial h +\overline{\partial h} =
\sum_{k=1}^\infty 2k{\rm Re}(h_k\zeta^{k-1}), \nn \\
a = 2\sum_{k=1}^\infty {\rm Re}(h_k\zeta^{k-1})A_k(\zeta\bar\zeta),
\nn \\
{\cal D}a = 2\sum_{k=1}^\infty {\rm Re}(h_k\zeta^{k-1})
\Big((k-1)A_k(\zeta\bar\zeta)
+ 2\zeta\bar\zeta A'_k(\zeta\bar\zeta)\Big),
\nn \\
{\cal D}a -a = 2\sum_{k=1}^\infty {\rm Re}(h_k\zeta^{k-1})
\Big((k-2)A_k(\zeta\bar\zeta)
+ 2\zeta\bar\zeta A'_k(\zeta\bar\zeta)\Big),
\nn \\
\partial a =
\sum_{k=2}^\infty (k-1)h_k\zeta^{k-2}A_k(\zeta\bar\zeta) +
2\zeta\bar\zeta \sum_{k=1}^\infty {\rm Re}(h_k\zeta^{k-2})
A'_k(\zeta\bar\zeta) \ee (note that there is no singular term with
$\zeta^{-1}$ in the last line) and perform angular integration. Then
the term linear in $h$ vanishes (it is proportional to $h_1$ and
$\bar h_1$ which we put equal to zero), and the $h^2$-term in the
action is proportional to:
$$
\sum_{k=2}^\infty |h_k|^2 \int_0^1 \frac{\rho
d\rho}{(1-\rho^2)^{3/2}} \left\{
\rho^{2(k-2)}\Big(\frac{(k-1)^2}{2}A_k^2+(k-1)\rho^2A_kA'_k
+\rho^4(A'_k)^2\Big) -
\frac{1}{4}\rho^{2(k-1)}\Big((k-2)A_k+2\rho^2A'_k\Big)^2 +\right.
$$ $$
+ k^2\rho^{2(k-1)}\Big(1-\frac{1}{2}\rho^2\Big)
- \frac{1}{4}k^2\rho^4\rho^{2(k-1)}
+\Big(1+\frac{1}{2}\rho^{2})k\rho^{2(k-1)}A_k
- \frac{1}{2}k \rho^{2k} \Big((k-1)A_k + 2\rho^2A'_k\Big)-
$$
\be
\left.
+3\frac{1}{1-\rho^2}\left(\frac{1}{2}\rho^{2(k-1)}A_k
\Big((k-2)A_k+2\rho^2A'_k\Big)
- (1-\frac{1}{2}\rho^2)\rho^{2(k-1)}kA_k
\right)
+\frac{15}{4(1-\rho^2)^2}\rho^{2(k-1)}A_k^2
\right\}
\label{bulkin}
\ee
The terms independent on $A_k$ and their derivatives are collected into
$\sigma_k^{(0)}$, those linear in $A_k$ and their derivatives into
$\sigma_k^{(1)}$, and the quadratic terms into $\sigma_k^{(2)}$
\be
(\ref{bulkin})=\sigma_k^{(0)}+\sigma_k^{(1)}+\sigma_k^{(2)}
\ee

\subsection{Calculating integrals}

Now we calculate the integral (\ref{bulkin}). To this end, note that
since $F_k$'s satisfy the equations
\be
x(1-x)F''_k(x) + \left(k+1 - \Big(k+\frac{1}{2}\Big)x\right)F'_k(x)
- \frac{k(k-1)}{4}F_k(x) = 0
\ee
the functions $A_k(x) = \frac{F_{k-1}(x)}{F_{k-1}(1)} - x$
satisfy
\be
x(1-x)A''_k + \left(k - \Big(k-\frac{1}{2}\Big)x\right)A'_k -
\frac{(k-1)(k-2)}{4} A_k = 
\frac{k(k+1)}{4}x - k
\label{eqforA}
\ee
The terms
\be
\sigma_k^{(0)}=\int\frac{dx}{(1-x)^{3/2}}
k^2x^{k-1}\left(1-\frac{x}{2}x-\frac{x^2}{4} \right)
\ee
and
\be
\sigma_k^{(1)}=
\int\frac{dx}{(1-x)^{3/2}}\left\{-kx^{k+1}A'_k +
x^{k-1}\left(k\left(1+\frac{1}{2}x\right)
-\frac{1}{2}k(k-1)x - \frac{3(2-x)k}{2(1-x)} \right)A_k\right\}
\ee
are immediately\footnote{
Since
\be
-\int\frac{dx}{(1-x)^{3/2}}kx^{k+1}A'_k =
-\lim_{x\rightarrow 1-0}\frac{kA_k x^{k+1}}{(1-x)^{3/2}}
+ \int\frac{A_kx^kdx}{(1-x)^{3/2}}
\Big(k(k+1)+\frac{3kx}{2(1-x)}\Big),
\ee
} reduced to (\ref{sigmas}), while in order to calculate
\be
\sigma_k^{(2)}=\int\frac{dx}{(1-x)^{3/2}}\left\{x^k(1-x)(A'_k)^2 +
2 x^{k-1}\left(\frac{k-1}{2} - \frac{(k-2)x}{2}
+\frac{3x}{2(1-x)}\right)A_kA'_k + \right.
\\
\left.
+  x^{k-2}\left(\frac{(k-1)^2}{2} - \frac{(k-2)^2x}{4}
+\frac{3(k-2)x}{2(1-x)} + \frac{15x}{4(1-x)^2}
\right)A_k^2 -\right.
\label{intwA}
\ee
we integrate the first term by parts and make use of
(\ref{eqforA}):
$$
\int\frac{dx}{(1-x)^{3/2}}x^k(1-x)(A'_k)^2 =
\lim_{x\rightarrow 1-0} \frac{x^kA_kA'_k}{\sqrt{1-x}}
- \int\frac{x^{k-1}dx}{(1-x)^{3/2}}\left(x(1-x)A''_k +
\left((1-x)k+\frac{x}{2}\right)A_k'\right)A_k =
$$
\be
= \lim_{x\rightarrow 1-0} \frac{x^kA_kA'_k}{\sqrt{1-x}}
- \frac{1}{4}\int\frac{x^{k-1}dx}{(1-x)^{3/2}}
\Big((k-1)(k-2)A_k^2 + \big(k(k+1)x - 4k\big)A_k\Big)
\ee
Integrating by parts the second term in (\ref{intwA})
we obtain:
$$
\int\frac{x^{k-1}dx}{(1-x)^{3/2}}
\left(\frac{k-1}{2} - \frac{(k-2)x}{2}
+\frac{3x}{2(1-x)}\right)(2A_kA'_k) =
\lim_{x\rightarrow 1-0}\frac{x^{k-1}A_k^2}{(1-x)^{3/2}}
\left(\frac{k-1}{2} - \frac{(k-2)x}{2}
+\frac{3x}{2(1-x)}\right) -
$$
$$
-\int\frac{x^{k-2}dx}{(1-x)^{3/2}}
\left(\frac{(k-1)^2}{2}-\frac{k(k-2)x}{2} + \frac{3kx}{2(1-x)}
+ \frac{3x}{2(1-x)}\Big(\frac{k-1}{2}-\frac{(k-2)x}{2}\Big)
+ \frac{15x^2}{4(1-x)^2}\right)A_k^2
$$
Collecting all the terms with $A_k^2$,
$$
 \int\frac{A_k^2x^{k-2}dx}{(1-x)^{3/2}}\left\{\left(
\frac{(k-1)^2}{2} - \frac{(k-2)^2x}{4}
+\frac{3(k-2)x}{2(1-x)} + \frac{15x}{4(1-x)^2}\right)
-\frac{(k-1)(k-2)x}{4} -\right.
$$ $$
\left.-\left(\frac{(k-1)^2}{2}-\frac{k(k-2)x}{2}
+ \frac{3kx}{2(1-x)}
+ \frac{3x}{2(1-x)}\Big(\frac{k-1}{2}-\frac{(k-2)x}{2}\Big)
+ \frac{15x^2}{4(1-x)^2}
\right)\right\} =
$$
$$
=  \int\frac{A_k^2x^{k-1}dx}{(1-x)^{3/2}}\left\{
\frac{k-2}{4}\Big(-(k-2) - (k-1) +2k\Big) +
\frac{3}{2(1-x)}\left(k-2 + \frac{5}{2} -k -
\Big(\frac{k-1}{2}-\frac{(k-2)x}{2}\Big)\right)
\right\} =
$$
\be =  \int\frac{A_k^2 x^{k-1}dx}{(1-x)^{3/2}}\left\{
\frac{3(k-2)}{4} + \frac{3}{4(1-x)}\Big(1-(k-1)+(k-2)x\Big) \right\}
= 0 \label{A2terms} \ee and one remains only with \be
\lim_{x\rightarrow 1-0} \frac{x^kA_kA'_k}{\sqrt{1-x}} -
\frac{1}{4}\int\frac{x^{k-1}dx}{(1-x)^{3/2}} \big(k(k+1)x -
4k\big)A_k \ee which is the same as $\sigma_k^{(2)}$ in
(\ref{sigmas}), since the boundary term vanishes.

\section{Conformal symmetry
\label{cosy}}
\setcounter{equation}{0}

In our $ADS_3$-restricted problem the global
conformal symmetry of \cite{dks,dhks1,dhks2,dhks3,koma} reduces to
$SL(2)$ with three complex-valued generators.
In what follows we use the formulation of \cite{koma}.

\subsection{$SL(2)$ action at the boundary}

When acting on a functional $F\{z(s)\}$ of parameterized
curve $\Pi: S^1 \rightarrow C$, the three generators are
\be
\hat J_{-} F = \oint  \frac{\delta F}{\delta z(s)}ds, \nn \\
\hat J_0 F = \oint  z\frac{\delta F}{\delta z(s)}ds, \nn \\
\hat J_{+} F = \oint  z^2\frac{\delta F}{\delta z(s)}ds
\label{Jgen}
\ee
There are additional three complex-conjugate operators.
Since in \cite{koma} the general situation (beyond complex plane)
is considered, the third generator in (\ref{Jgen})
was written in a more general form
\be
\hat{{\vec {\cal J}}_{-}}= \oint ds \frac{\delta }{\delta \vec y(s)},\nn \\
\hat{{{\cal J}}_{0}}  = \oint ds
\left(\vec y(s)\frac{\delta }{\delta \vec y(s)}\right), \nn \\
\hat{{\vec {\cal J}}_{+}}  = \oint ds \left\{2\,\vec y(s)
\left(\vec y(s)\frac{\delta }{\delta \vec y(s)}\right) -
\vec y\,^2(s)\frac{\delta }{\delta \vec y(s)}\right\}
\ee
They are ${\vec {\cal J}}_-=\left(J_{-},\bar J_{-}\right)$,
${\cal J}_0=J_{0}+\bar J_{0}$ and
${\vec {\cal J}}_+=\left(J_{+},\bar J_{+}\right)$ in our situation.

We now need to express these generators in terms of $h_k$
variables.
From $z = \zeta + \sum_k h_k \zeta^k$,
$\bar z = \bar\zeta + \sum_k \bar h_k \bar\zeta^k$ and
\be
\delta F = \oint\frac{\delta F}{\delta z(s)}\delta z(s)ds +
\oint\frac{\delta F}{\delta \bar z(s)}\delta \bar z(s)ds =
\sum_k \delta h_k \oint \frac{\delta F}{\delta z(s)}\zeta^k(s) ds
+ \sum_k \delta\bar h_k
 \oint \frac{\delta F}{\delta \bar z(s)}\bar \zeta^k(s) ds
\ee
we conclude that
\be
\oint \frac{\delta F}{\delta z(s)}\zeta^k(s) ds =
\frac{\partial F}{\partial h_k}, \ \ \ \
\oint \frac{\delta F}{\delta \bar z(s)}\bar\zeta^k(s) ds =
\frac{\partial F}{\partial \bar h_k}
\ee
Therefore
\be\label{cs}
\hat J_{-}  = \frac{\partial}{\partial h_0}, \nn \\
\hat J_0 = \frac{\partial}{\partial h_1} + \sum_{k=0}^\infty h_k
\frac{\partial}{\partial h_k}, \nn \\
\hat J_{+} = \frac{\partial}{\partial h_2} + 2\sum_{k=0}^\infty
h_k \frac{\partial}{\partial h_{k+1}} +
\sum_{k,l=0}^\infty h_kh_l\frac{\partial}{\partial h_{k+l}}
\ee

\subsection{Invariance properties of $h$-series: a surprise}

It is easy to check that (\ref{Dans}) is invariant
under these $SL(2)$ transformations, while (\ref{Aans}) is not.
Indeed, $\hat J_-$ annihilates all $h$-series that do
not contain $h_0$ -- and both (\ref{Dans}) and (\ref{Aans})
belong to this class.

The relevant properties of the coefficients in
(\ref{Dans}) are:
\be
C_{1k} = B_k,\ \ \   C_{11}=C_{12} = 0, \ \ \
C_{2k} = 2B_{k+1},\ \ \
V_{1kl} = 2C_{kl}, \ \ \
V_{2kl}+2A_{k+l}=2\left(C_{k,l+1}+C_{k+1,l}\right)\\
U_{ij;1l}=C_{ij},\ \ \ U_{ij;2l} = 2C_{ij}
\ee
They are indeed satisfied by the coefficients
$B^{(D)}$, $C^{(D)}$, $V^{(D)}$ and $U^{(D)}$ in (\ref{Dans}).
At the same time for (\ref{Aans})
$C_{22}^A \neq 2B_3^A$!

This fact is somewhat surprising because one could expect the opposite result:
the double integral $D_\Pi$ is not {\it a priori}
annihilated by $\hat J_{+1}$, while $A_\Pi$ is shown to be
invariant \cite{koma}. In particular, the BDS formula is known to satisfy
(anomalous) conformal Ward identities for all $\Pi$
\cite{dks,dhks1,dhks2,dhks3,koma}.
Indeed, dilogarithms in the BDS formula \cite{bds}
depend only on invariant cross-ratios,
while logarithms reproduce the anomaly part of the Ward identity.

\subsection{Invariance of the double integral}

To explain the invariance of the double integral, one should note that
the integrand in $D_\Pi$ is obviously not invariant under the projective
transformations generated by (\ref{Jgen}), instead it
changes by a total derivative. Therefore, as soon as the integral diverges,
one has to be careful with its invariance. Indeed, one can easily see
the divergent part is not projective-invariant: it is proportional to
the curves length
$L=\oint dl = \oint \sqrt{\dot z\dot{\bar z}}ds$,
which transforms as follows:
\be\label{sV}
\hat J_{-} L = 0, \nn \\
\hat J_{0} L = {L\over 2}, \nn \\
\hat J_{+} L = \oint z dl
\ee
as can be read off from formulae (\ref{Jgen}).

At the same time, this quite formal calculation can be confirmed from
explicit manipulations with the $h$-series. When $J_0$ from (\ref{cs})
acts on $L$ which is given by
formula (\ref{lenc}) with $h_1$ and $\bar h_1$ switched on, then it converts the
typical term in the $h$ series for $L$,
\be
\sqrt{(1+h_1)(1+\bar h_1)}\left({h\over 1+h_1}\right)^p
\left({\bar h\over 1+\bar h_1}\right)^q
\ee
into
\be
\hat J_0 \sqrt{(1+h_1)(1+\bar h_1)}\left({h\over 1+h_1}\right)^p
\left({\bar h\over 1+\bar h_1}\right)^q=\left[p-(p-{1\over 2})\right]
\sqrt{(1+h_1)(1+\bar h_1)}\left({h\over 1+h_1}\right)^p
\left({\bar h\over 1+\bar h_1}\right)^q
\ee
i.e.
\be
\hat J_0 L={L\over 2}
\ee
as required in (\ref{sV}).

Similarly, one can use the explicit form of $\hat J_+$ in terms of $h$,
\be
\hat J_+=h_0^2{\partial\over\partial h_0}+2h_0\hat J_0+(1+h_1)^2{\partial\over
\partial h_2}+2(1+h_1)\sum_{k=2}h_k{\partial\over\partial h_{k+1}}+
\sum_{k,l\ge 2}h_kh_l{\partial\over h_{k+l}}
\ee
and act with it on $L$ from (\ref{lenc}),
\be
{1\over 2\pi}\oint dl=\left|1+h_1\right| + \frac{1}{4}\sum_{k=2} {k^2
|h_k|^2\over \left|1+h_1\right|} - \frac{1}{16}\sum_{k,l=2}^\infty
kl(k+l-1) \left[{h_kh_l\bar h_{k+l-1}\over
\left(1+h_1\right)\left|1+h_1\right|} + {\bar h_k\bar h_l
h_{k+l-1}\over \left(1+\bar h_1\right)\left|1+h_1\right|}\right]+
\ldots
\ee
to obtain
\be
\hat J_+ {L\over 2\pi}=2h_0\left(\hat J_0{L\over 2\pi}\right)+(1+h_1)^2\left[{2^2\over 4}\bar h_2-{1\over 16}
{2\cdot 2\over (1+h_1)|1+h_1|}\sum_{k=2}k(k+1)h_k\bar h_{k+1}\right]+\\+
2(1+h_1){1\over 4|1+h_1|}\sum_{k=2}(k+1)^2h_k\bar h_{k+1}=\\
={h_0\over 2\pi}\oint dl+(1+h_1)^2{\bar h_2\over |1+h_1|}+\sum_{k=2}{(k+1)(k+2)\over 4}
{1+h_1\over |1+h_1|}h_k\bar h_{k+1}={1\over 2\pi}\oint zdl
\ee
in accordance with (\ref{sV}). Note that this calculation depends on the
explicit form of $h^3$-terms.

\subsection{On symmetries of the minimal area}

First of all, the r.h.s. of the anomalous Ward identity
(A.19) in \cite{koma} vanishes in our smooth $n=\infty$ limit.
Therefore, according to \cite{koma} the minimal area is conformal invariant!
-- what seems to contradict apparent non-invariance of $A_\Pi$.

For an {\it a priori} check of the symmetry of the minimal action
one needs to extend the action of $SL(2)$ from the boundary
to entire $AdS$ space. The group action is \cite{koma}:
\be
r \rightarrow \frac{r}{1+2\vec\beta\vec y + \vec\beta^2(r^2+\vec y^2)},
\nn \\
\vec y \rightarrow \frac{\vec y + \vec\beta(r^2+\vec y^2)}
{1+2\vec\beta\vec y + \vec\beta^2(r^2+\vec y^2)} \ee At the boundary
$r^2=0$ it reduces to the projective transformation $z\to
z+\bar\beta z^2+O(\beta)$. The problem is that for $r^2\neq 0$ the
action of $\hat J_+$ on $z$ transforms it into non-holomorphic
function of $\zeta$. Application of the Gauss-Riemann decomposition
is needed to restore holomorphicity, what can imply a more
sophisticated action on $h$-variables beyond the boundary. It can
happen that such modifications involve $\mu$-linear terms, which can
generate $\mu$-finite corrections from the variation of $L/\mu$
contributions. This is also a kind of anomaly -- which needs to be
studied more accurately. This anomaly in conformal symmetry
(\ref{cs}) is a part of a larger anomaly for $n=\infty$ discovered in this paper,
which, in its turn, generalizes the Alday-Maldacena result,
\cite{am3}. A similar anomaly for $n=6$ was recently found in \cite{dhks3}, see also
a fresh additional evidence in \cite{bdkrsvv,dhks4}.

\section*{Acknowledgements}

We are indebted for short but stimulating discussions of
Alday-Maldacena duality  at $n=\infty$ to
Katsushi Ito, Antal Jevicki,
Alexander Gorsky, Hikaru Kawai, Horatiu Nastase,
Arkady Tseytlin, Anton Zabrodin and Konstantin Zarembo.
Our understanding of entire subject was very much
affected by common work with Theodore Tomaras.

The work of A.M.'s is partly supported by Russian Federal Nuclear
Energy Agency, by the joint grant 06-01-92059-CE,  by NWO project
047.011.2004.026, by INTAS grant 05-1000008-7865, by
ANR-05-BLAN-0029-01 project and by the Russian President's Grant of
Support for the Scientific Schools NSh-3035.2008.2, by RFBR grants
07-02-00878 (A.Mir.) and 07-02-00645 (A.Mor.).

\newpage

\section*{Appendix I: $h$-series representation of
reparametrization invariants at the boundary
\label{invs}}
\setcounter{equation}{0}
\def\theequation{A.\arabic{equation}}

Here we express the circumference of the deformed circle and the integral
of logarithm of its curvature as functions of coefficients $h_k$ up to the
second order in these coefficients (we still put $h_1=\bar h_1=0$).

With the conformal map
\be
z = e^{i\varphi} + \sum_{k=0}^\infty h_ke^{ik\varphi}
\ee
the square of length element is
\be
dl^2 = \left|\frac{dz}{d\varphi}\right|^2 =
\left|1 + \sum kh_k e^{i(k-1)\varphi}\right|^2
= 1 + 2\sum_{k=1}^\infty {\rm Re} \left(kh_ke^{i(k-1)\varphi}\right)
+ \sum_{k,l=1}^\infty kl{\rm Re} h_k\bar h_l e^{i(k-l)\varphi}
\label{dl2sum1}
\ee
Integration along the circle over $\frac{d\varphi}{2\pi}$
converts the sums of exponentials in the following way:
\be
\begin{array}{ccc}
\sum_{k=2}^\infty f(k) {\rm Re}\left(h_ke^{i(k-1)\varphi}\right)
& \longrightarrow & 0 \\
\sum_{k,l=1}^\infty {\rm Re} f(k,l) h_k\bar h_l e^{i(k-l)\varphi}
& \longrightarrow & \sum_{k=1}^\infty f(k,k)|h_k|^2 \\
\left\{
\sum_{k=1}^\infty f(k) {\rm Re} \left(h_ke^{i(k-1)\phi}\right)\right\}
\left\{
\sum_{k=1}^\infty g(k) {\rm Re} \left(h_ke^{i(k-1)\phi}\right)\right\}
& \longrightarrow & \frac{1}{2}\sum_{k=1}^\infty f(k)g(k)|h_k|^2
\end{array}
\label{averangles1}
\ee
Keeping this in mind, one gets
\be
\frac{1}{2\pi}L
= \int dl = 1 + \frac{1}{4}\sum_{k=1}^\infty k^2|h_k|^2
+ O(h^3)
\ee

Proceed now to curvature and its derivatives.
The first local reparametrization invariant (scalar) of a curve
is its scalar curvature,
\be
\kappa
= \frac{{\rm Im} \Big(\ddot z \dot{\bar z}\Big)}
{ |\dot z \dot {\bar z}|^{3/2} }
\ee
and
\be
\frac{1}{2\pi}\int \log\kappa \,dl
= -\frac{1}{4}\sum_{k=1}^\infty (k^2-4k+5)k^2|h_k|^2
\ee
Similarly, one can calculate $d\kappa/d\varphi$, $\dot\kappa\equiv d\kappa/dl$
and the integral of square of this latter, the result reads
\be
\frac{1}{2\pi}\int \left(\frac{d\kappa}{dl}\right)^2 \,dl
= \frac{1}{2}\sum_{k=3}^\infty \Big(k(k-1)(k-2)\Big)^2|h_k|^2
\ee

\section*{Appendix II: An alternative calculation of $D_{\Pi}$
by $r^{\prime}/r$ regularization}

     In this appendix,  we compute the double contour integral ${D}_\Pi$
       to the quadratic order,
     regularizing the integral by making  the relative size of  the two radii $r, r^{\prime}$
     associated with the two circular line integrals  in $\zeta$ plane  different from unity.
     This is a version of $c$-regularization,
     an alternative regularization to the ``$\lambda$" regularization in the text.
   Let $r r^{\prime} =1$, $z= H(\zeta) = \zeta + h(\zeta)$.
 \be
   {D}_{\Pi} =  \oint_{\Pi_{r}} \oint_{\Pi_{r^{\prime}}}  \frac{\frac{1}{2}(dz d \bar{z}^{\prime}
    + d \bar{z} dz^{\prime}) }{(z-z^{\prime})(\bar{z}-\bar{z}^{\prime})}
         =  {D}_{\Pi}^{(0)} +   {D}_{\Pi}^{(1)} +  {D}_{\Pi}^{(2)}  + O(h^3)  ,
 \ee
     where  ${D}_{\Pi}^{(i)}, i= 0,1,2$ denote order $h^0$, $h^1$ and $h^2$ contribution to
       ${D}_{\Pi}$ respectively. It is immediate to see that ${D}_{\Pi}^{(1)} = 0$ and
 \be
   {D}_{\Pi}^{(0)} = \oint_{ |\zeta| =r} \oint_{ |\zeta^{\prime}| = r^{\prime}}
    \frac{\frac{1}{2}(d \zeta d \bar{\zeta}^{\prime}
    + d \bar{\zeta} d{\zeta}^{\prime}) }{(\zeta - \zeta^{\prime})(\bar{\zeta}-\bar{\zeta}^{\prime})}  .
\label{dpi0}
 \ee
    After splitting the double integral into that over total and relative angles,
    $\Phi$ and $\varphi$,  (\ref{dpi0})
     becomes  a simple Poisson integral:
 \be
   2(2\pi) a \int_{-\pi}^{\pi}  \frac{\cos \varphi d \varphi}{1- 2a \cos \varphi + a^2}
       =  2(2\pi)^2 \left( \frac{1}{1-a^2} - 1 \right).
 \ee
     where $a \equiv  \frac{r^{\prime}}{r}  $.
   As for ${D}_{\Pi}^{(2)}$,  after some  calculation,  we obtain
 \be
   {D}_{\Pi}^{(2)}   = \sum _{k, \ell} h_{k} \bar{h}_{\ell} \left[  \frac{1}{2}
     \oint \oint d \zeta d \bar{\zeta}^{\prime}
       \frac{ \zeta^{k-1} {\bar{\zeta}}^{\prime \ell-1} f_{k}(\zeta^{\prime}/\zeta)
         f_{\ell}(\bar{\zeta}/\bar{\zeta}^{\prime}) }{ (\zeta -\zeta^{\prime})(\bar{\zeta}
          - {\bar{\zeta}}^{\prime})}
          + \frac{1}{2}
     \oint \oint d \bar{\zeta} d{\zeta}^{\prime}
      \frac{ {\zeta^{\prime}}^{k-1} {\bar{\zeta}}^{\ell-1} f_{k}(\zeta/\zeta^{\prime})
         f_{\ell}(\bar{\zeta}^{\prime}/\bar{\zeta}) }{ (\zeta -\zeta^{\prime})(\bar{\zeta}
          - \bar{\zeta}^{\prime})}  \right],
 \ee
     where
 \be
 f_{k}(x) =  \frac{1-x^k}{1-x} -k.
 \ee
We put $a=1$, since
   the integral is finite at this point.  Making a change of variables
   $\zeta = e^{i(\Phi - \frac{1}{2} \varphi)},
    \zeta^{\prime} = e^{i(\Phi + \frac{1}{2} \varphi)} ,  w= e^{i\varphi}$, and
      carrying out the $d\Phi$ integral,  we obtain
 \be
   - (2\pi)^{2} \sum_{k} |h_{k}|^2 \left[  \oint \frac{dw}{2\pi i} \frac{w^{-k} f_{k}(w)^2}{(1-w)^2}
      + c.c. \right]
      =     -2 (2\pi)^2 \sum_{k} |h_{k}|^2 \left( \sum_{i=0}^{k-1} c_{i}c_{k-1-i} \right)
       = -2 (2\pi)^2  Q_{\Pi}^{(2)},
 \label{quadD}
 \ee
 where  $c_{\ell}$ with $c_{i}= -(k-1) +i,$  for  $ 0 \leq i \leq k-1$ are the Taylor coefficients
 \be
  \frac{f_{k}(x)}{(1-x)}   =  \sum_{n=0}^{\infty} c_{n}x^n .
 \ee
   Eq. (\ref{quadD}) agrees with  the result (\ref{Dans}) of calculations in
   the $\lambda$ regularization.

 To summarize,
 \be
 {D}_{\Pi} =  2(2\pi)^2 \left( \frac{1}{1-a^2} - 1 \right)  -2 (2\pi)^2  Q_{\Pi}^{(2)}+ O(h^3).
 \ee
   Higher order computation can be carried out as is in the main text.

\section*{Appendix III: Circle vs. rectangular}

In this Appendix, we comment on technical differences between
the long rectangular that was considered in \cite{am3} and the deformed
circle we consider in the paper.

\subsection*{Asymptotic behavior of $r$ near the boundary}

First of all, let us consider the behaviour of solution to the NG equation.
From (\ref{NGeqninf1}) in the leading order in
$y_\bot$ and $y_{||}$ we get
\be
r = \sqrt{\frac{2y_\bot - \kappa y_{||}^2}{\kappa}}
\ee
where $\kappa$ is the curvature
(inverse radius of the tangent circle)
at the given point of the boundary.
This can be considered as a limit near the boundary
of exact circle solution (\ref{NGcirc}),
\be
r = \sqrt{\kappa^{-2} - (\kappa^{-1}-y_\bot)^2 - y_{||}^2}
\label{asyr}
\ee
Technically the contribution to (\ref{NGeqninf1})
in this order comes from
\be
\partial_\bot r = \frac{1}{\kappa r}, \nn \\
\partial^2_\bot r = -\frac{1}{\kappa^2 r^3}, \nn \\
\partial_{||}^2 r = -\frac{1}{r} + O(y_{||}^2)
\ee
the first derivative $\partial_{||} r$ and the mixed derivative
$\partial_\bot\partial_{||} r$ are proportional to $y_{||}$ and
can be neglected.
Then the relevant terms in (\ref{NGeqninf1}) are
\be
2(\partial_\bot r)^2 + r\partial^2_\bot r
+ r(\partial_\bot r)^2\partial^2r
= r(\partial_\bot r)^2\partial^2_\bot r
\ee
Since at $r\rightarrow 0$ the second derivative
$\partial_{||}^2 r \ll \partial_{\bot}^2 r$, it
can be neglected in the $r\partial^2 r$ term, but it contributes
to the $r^4$ terms, because it is multiplied by a large
factor $\partial_\bot r$. These two $r^4$ terms actually
combine into $r(\partial_\bot r)^2\partial^2_{||} r$ at the l.h.s.
and this contribution is crucial for (\ref{asyr}) to be
a solution to (\ref{NGeqninf1}): the three terms at the l.h.s.
contribute $2-1-1=0$.

Already from this calculus it is clear that things will go wrong if
(\ref{asyr}) does not depend on $y_{||}$. This happens when the
boundary straightens, $\kappa = 0$, even at a single point - nothing
to say about the boundary containing entire straight segments like
in \cite{am3}.\footnote{ It deserves emphasizing that we speak here
about a straight segment in {\it projection} $\bar \Pi$ in the
$n=\infty$ limit: this argument is non-applicable neither to the
light-like straight segments which compose $\Pi$, nor to the
finite-$n$ polygons, where $\bar\Pi$ consists of straight segments,
but $y_0$ can not be neglected, as in \cite{am1,mmt1,imm}.} The
problem is already seen in (\ref{asyr}): $\kappa$ enters also as a
normalization factor and stands in the denominator. Clearly, at
$\kappa=0$ asymptotics (\ref{asyr}) is seriously modified, actually
it is substituted by \be r \sim \sqrt[3]{y_\bot}, \label{asyr3} \ee
(note that $\sqrt[3]{y_\bot} \gg \sqrt{y_\bot}$ at small $y_\bot$).
The interpolating formula \be 2y_\bot - \kappa y_{||}^2 = \kappa r^2
+ \hbox{const}\cdot r^3 + O(r^4) \ee

The situation gets even more tricky if convexity of the curve
$\bar\Pi$ is changed: solution (\ref{asyr}) turns imaginary
at the other side of the boundary -- i.e. simply fails to exist.
This means that, near the boundary, the minimal surface is locally bent
towards the center of curvature of the boundary.

In any case we see that at $n=\infty$ the $\bar\Pi$ with some
straight segments is a kind of a very special limit, considerably
different from generic situation.
This can imply that the long-rectangular example of \cite{am3},
despite its seeming simplicity can actually be non-trivial and
require a more serious analysis.
We, however, restrict ourselves to a brief reminder of that
example in the next subsection.

\subsection*{An example of rectangular}

We calculate here the minimal area of the rectangular and
demonstrate it does not look like the double contour integral
\cite{MR}.

We consider a very long rectangular of the length $L_\parallel$ and the
width $L$ so that the solution to the
NG equations depends on the only perpendicular variable $y_\perp=y$.
Then, the solution $r(y)$ is easier written in terms of the inverse function
$$
y(r)=\int_0^r{\xi^2d\xi \over\sqrt{C^4-\xi^4}}=-CD\left(\arcsin{r\over
C},i\right)
$$
for $0\le y\le L/2$ and the opposite sign of the root for $L/2\le
y\le L$. $D(x,k)\equiv F(x,k)-E(x,k)$ here is the difference of
elliptic integrals of the first and the second kinds respectively.
Then,
$$
L=2y(C)=2\sqrt{2}C\left(E-{K\over 2}\right)=2\sqrt{2}C{\pi\over 4K}
$$
where $E$ and $K$ are complete elliptic integrals of the first and
the second kinds respectively taken at the value of elliptic modulus
$k=k'=1/\sqrt{2}$. Note that in this lemniscata point $K={\Gamma
(1/4)^2\over 4\sqrt{\pi}}$ and, using the Legendre formula
$$
KE'+K'E-KK'={\pi \over 2}
$$
for the four complete elliptic integrals with complimentary
modulus, one immediately obtains $E={\pi\over 4K}+{K\over 2}$. Then,
one obtains
$$
L={\pi C\over \sqrt{2}K}, \ \ \ \ \hbox{i.e.}\ \ \
C={\sqrt{2}KL\over\pi}
$$

The area is ($\mu^2$ is the regulator)
$$
S=2L_\parallel
C^2\int_0^C{dr\over (r^2+\mu^2)\sqrt{C^4-r^4}}={2L_\parallel\over C}\int_0^1
{dr\over (r^2+\mu^2)\sqrt{1-r^4}}={\sqrt{2}L_\parallel\over C}{1\over
1+\mu^2}\Pi\left(-{1\over 1+\mu^2},{1\over\sqrt{2}}\right)
$$
where $\Pi(\nu,k)$ is the complete elliptic integral of the third
kind. Its asymptotics can be found from the relation
$$
k'^2{\sin\theta\cos\theta\over 1-k'^2\sin^2\theta}
\Big[\Pi\Big(-(1-k'^2\sin^2\theta),k\Big)-K\Big]={\pi\over
2}-(E-K)F(\theta,k')-KE(\theta,k)
$$
and using $F(\theta,k)=\theta+O(\theta^3)$,
$E(\theta,k)=\theta+O(\theta^3)$:
$$
\Pi\left(-{1\over
1+\mu^2},{1\over\sqrt{2}}\right)={\pi\over\sqrt{2}\mu}-{\pi\over
2K}+O(\mu)
$$

Then, the area
$$
S={\pi L_\parallel\over C}\left({1\over\mu}-{1\over\sqrt{2}K}\right)=
{\pi^2L_\parallel\over\sqrt{2}KL}\left({1\over\mu}-{1\over\sqrt{2}K}\right)
$$
The finite piece in this answer is
\be\label{AMd}
S_{fin}=-{\pi^2L_\parallel\over 2KL}=-{(2\pi)^3\over \Gamma(1/4)^4}
{L_\parallel\over L}
\ee
This result has to be compared with the double contour integral. Its finite part
comes from the case when $y$ and $y'$ belong to two different parallel lines
(when they belong to the same line one gets the contribution to the divergent
term)
\be\label{dlilore}
2L_\parallel\int_{-\infty}^{+\infty}{d\xi\over \xi^2+L^2}=2\pi {L_\parallel
\over L}
\ee
The difference between $2\pi$ in (\ref{dlilore}) and
the coefficient in (\ref{AMd}) is the confusing problem
discovered in \cite{am3}. One can formulate our result as a non-trivial
generalization of this statement:

$\bullet$ A {\it similar} coefficient discrepancy exists for a circle of arbitrary
shape, but only {\it few} of infinitely many coefficients are
{\it different}.

\section*{Appendix IV: MAPLE programs
\label{progs}}
\setcounter{equation}{0}

We append here two simple MAPLE programs that one can use for
evaluating the minimal area and the double contour integral (the
latter one up to any given order in $h$). Using this program to
obtain the area up to $h^4$ order and higher requires the knowledge
of solution to the NG equation up to this order.

\subsection*{Calculation of $A_\Pi$}

Literally, this program calculates the finite part $CCfin$ of the coefficient
in front of the cubic term $h_kh_l\bar h_{k+l-1}$. It uses the explicit form (\ref{athrh})
of the NG-harmonic functions.

\begin{verbatim}
>dH:=z->1+s*dh(z): dHH:=z->1+s*dhh(z):r2:=1-z*zz+s*a(z,zz):
>
> S:=sqrt( dH(z)*dHH(zz)*(dH(z)*dHH(zz)*r2 + diff(r2,z)*diff(r2,zz)) )/r2^(1/2)/(r2+mu^2);
>
> SS:=mtaylor(simplify(mtaylor(S,s,1)*(1-z*zz+mu^2)^(3/2)),c,2);
> SL:=simplify(simplify(mtaylor(S,s,2)-mtaylor(S,s,1))*(1-z*zz+mu^2)^(5/2)/s);
> SQ:=simplify(simplify(mtaylor(S,s,3)-mtaylor(S,s,2))*(1-z*zz+mu^2)^(7/2)/s^2):
> SC:=simplify(simplify(mtaylor(S,s,4)-mtaylor(S,s,3))*(1-z*zz+mu^2)^(9/2)/s^2):
>
> A:=(k,z,zz)->(1+(k-1)*sqrt(1-z*zz))*(1-sqrt(1-z*zz))^(k-1)/(z*zz)^(k-1)-z*zz;
> K:=5: L:=5: M:=K+L-1:
> h:=z->h[K]*z^K+h[L]*z^L; hh:=z->hh[M]*z^(M);
>
> dh:=z->diff(h(z),z); dhh:=z->diff(hh(z),z);
> a:=(z,zz)-> h[K]*z^(K-1)*A(K,z,zz) + h[L]*z^(L-1)*A(L,z,zz) + hh[M]*zz^(M-1)*A(M,z,zz) ;
> diff(a(z,zz),z):
> SS1:=simplify(SS);
> SL1:=simplify(SL);
> SQ1:=simplify(SQ);
> SC1:=simplify(SC);
>
> z:=sqrt(X)*exp(I*phi): zz:=sqrt(X)*exp(-I*phi):
> #simplify(SL1);
> SLI:=factor(int(simplify(SL1),phi=0..2*Pi)/2/Pi);
> SQI:=factor(int(simplify(SQ1),phi=0..2*Pi)/2/Pi);
> SCI:=factor(int(simplify(SC1),phi=0..2*Pi)/2/Pi);
>
> LL:=factor(int(SLI/((1-X+mu^2)^(5/2)),X=0..1));
> QQ:=factor(int(SQI/((1-X+mu^2)^(7/2)),X=0..1));
> CC:=factor(int(SCI/((1-X+mu^2)^(9/2)),X=0..1));
>
> QQQ:=coeff(QQ,arctan(1/mu)); QQQQ:=subs(mu=0,simplify(QQ-QQQ*arctan(1/mu)));
> QQdiv:=coeff(simplify(QQQ*mu*(1+mu^2)^7),mu,0);
> QQfin:=simplify(QQQQ-QQdiv);
>
> CCC:=coeff(CC,arctan(1/mu)); CCCC:=subs(mu=0,simplify(CC-CCC*arctan(1/mu)));
> CCdiv:=coeff(simplify(CCC*mu*(1+mu^2)^11),mu,0);
> CCfin:=simplify(CCCC-CCdiv);

\end{verbatim}

\subsection*{Calculation of $D_\Pi$}

$N$ here denotes the number of switched on $h_k$, $NN\le k\le N$, and the calculation
is performed with the accuracy $O(h^N)$.

\begin{verbatim}
> p:=3:
> NN:=0: N:=6:
>
> # theta=Phi,  phi = varphi
>
> U:=exp(2*I*phi) + t*sum( (k*h[k]*exp(I*(k-1)*theta) +
> k*hh[k]*exp(-I*(k-1)*theta))*exp(I*(k+1)*phi), k=NN..N) +
> t^2*sum(sum(  k*l*h[k]*hh[l]*exp(I*(k-l)*theta)*exp(I*(k+l)*phi),
> l=NN..N),k=NN..N)  +  exp(-2*I*phi) +
> t*sum( (k*hh[k]*exp(-I*(k-1)*theta) +
> k*h[k]*exp(I*(k-1)*theta))*exp(-I*(k+1)*phi), k=NN..N) +
> t^2*sum(sum(  k*l*hh[k]*h[l]*exp(-I*(k-l)*theta)*exp(-I*(k+l)*phi),
> l=NN..N),k=NN..N);
>
> V:= simplify(1 + t*sum( simplify(sin(k*phi)/sin(phi))*(h[k]*exp(I*(k-1)*theta)+
> hh[k]*exp(-I*(k-1)*theta)), k=NN..N) +
> t^2*sum(sum( simplify(sin(k*phi)/sin(phi)*sin(l*phi)/sin(phi))*(h[k]*hh[l]*
> exp(I*(k-l)*theta)+hh[k]*h[l]*exp(-I*(k-l)*theta))/2, l=NN..N), k=NN..N));
>
> Ra:=(mtaylor(U/2/(4*sin(phi)^2*V + lambda^2),t,p+1));
> RA:=simplify(int( Ra, theta = 0..2*Pi)/2/Pi-1/(4*sin(phi)^2) + 1/2);
>
> DI:=simplify(int( RA, phi=0..2*Pi )/2/Pi);

\end{verbatim}

\end{document}